\documentstyle[12pt,leqno,twoside]{article}

\normalsize 
\divide\textheight by \baselineskip
\multiply\textheight by \baselineskip
\divide\oddsidemargin by 2
\include{mssymb12}
\input mssymb

\newtheorem{theorem}{Theorem}[section]
\newtheorem{corollary}{Corollary}[section]
\newtheorem{lemma}{Lemma}[section]
\newtheorem{conjecture}{Conjecture}[section]

\newtheorem{remark}{Remark}[section]
\newtheorem{definition}{Definition}[section]
\newtheorem{proposition}{Proposition}[section]

\newcommand{\binom}{\mbox{${\deg a} \choose i$}}
\newcommand{\binomhf}{\mbox{${\deg a -\hf} \choose i$}}

\newcommand{\cpn}{\mbox{$\tilde{\frak U}(\hat{\hat{\frak g}},k)
\langle z \rangle$}}
\newcommand{\hf}{\mbox{$\frac{1}{2}$}}
\newcommand{\thf}{\mbox{$\frac{3}{2}$}}
\newcommand{\hg}{\mbox{$\hat{\frak g}$}}
\newcommand{\hhg}{\mbox{$\hat{\hat{\frak g}}$}}
\newcommand{\hh}{\mbox{$\hat{\frak h}$}}
\newcommand{\hhn}{\mbox{$\hat{\hat{\frak n}}$}}

\newcommand{\hz}{\mbox{$\hf \Bbb Z$}}

\newcommand{\M}{\mbox{$M=\bigoplus_{n \in\frac{1}{2}\Bbb Z_{+}}M_n$}}

\newcommand{\strhf}{\mbox{$Res_z(Y(a,z)\frac {(z+1)^{\deg a-\hf}}{z}b)$}}

\begin{document}
\title{Vertex Operator Superalgebras and Their Representations
\footnote{This paper is in the final form, and no version of it will
be submitted for publication elsewhere.}}
\author{Victor Kac and Weiqiang Wang}
\date{}
\pagestyle{myheadings}
\markboth{Victor Kac and Weiqiang Wang}
{Vertex Operator Superalgebras and Their Representations}
\maketitle

\section*{0\,\,\,\,\,\,\,Introduction }

Vertex operator algebras (VOA)  were introduced
in physics by Belavin, Polyakov and Zamolodchikov [BPZ]
and in mathematics by Borcherds [B].
For a detailed exposition of the theory
of VOAs see [FLM] and [FHL]. In a remarkable development of
the theory, Zhu [Z] constructed
an associative algebra $A(V)$ corresponding to a VOA $V$ and
established a 1-1 correspondence
 between the irreducible representations of $V$ and
those of $A(V)$. Furthermore, Frenkel and Zhu [FZ]
defined an $A(V)$-module $A(M)$ for any $V$-module $M$
and then described the fusion rules in terms of the modules $A(M)$.
An important feature of these constructions is  that $A(V)$
and  $A(M)$ can usually
be computed explicitly. For example, they
enabled Frenkel and Zhu to prove the rationality and
compute the fusion rules
of VOAs associated to
the representations of affine Kac-Moody algebras
 with a positive integral level. They also allowed one of
the authors [W] to prove the rationality and
compute the fusion rules of VOAs associated to
the minimal series representations of the Virasoro algebra.
(Independently, Dong, Mason and Zhu [DMZ] proved the rationality
for the {\em unitary} minimal series of the  Virasoro algebra
and calculated
the fusion rules in the case of central charge $c= \hf$).

In this paper we generalize Frenkel-Zhu's construction
to vertex operator superalgebras (SVOA) and then
discuss in detail several interesting classes of SVOAs.
We present explicit formulas for the ``top'' singular vectors
and defining relations for the integrable representations
of the affine Kac-Moody superalgebras. These formulas
are not only crucial for the theory of the associated SVOAs
and their modules, but also of independent interest.

We organize this paper in the following way.
In Subsec.1.1 we present
definitions of vertex operator
superalgebras and their modules, emphasizing the existence
of the Neveu-Schwarz element in the so-called $N=1$ (NS-type)
SVOAs. We define in Subsec.1.2 an associative algebra $A(V)$
corresponding to a SVOA $V$ and
establish a bijective correspondence
between the irreducible representations of $V$  and
the irreducible representations of $A(V)$. In Subsec.1.3
we define an $A(V)$-module $A(M)$ for every $V$-module $M$ and then
describe the fusion rules in terms of modules $A(M)$. Needless to say
that, if we view a VOA as a SVOA with zero odd part,
then our construction reduces to Frenkel-Zhu's
original one.

In Subsec.2.1 we construct $N=1$
SVOAs $M_{k,0}$ and $L_{k,0}$
corresponding to the representations of an
affine Kac-Moody superalgebra $\hhg$.
In [KT], the minimal
representation $L(h^{\vee} \Lambda_0)$ of $\hhg$ was realized
in a Fock space $F$ of a certain infinite-dimensional
Clifford algebra contained in $\hhg$.
Kac and Todorov [KT] proved that
any unitary highest weight representation of $\hhg$
is of the form
$L(\Lambda + h^{\vee}\Lambda_0) = F \otimes \bar{L}(\Lambda)$, where
$\bar{L}(\Lambda)$ is the irreducible unitary
highest weight representation of
the affine Kac-Moody algebra $\hg$.
Explicit formulas for the ``top''
singular vectors of the Verma module $M(\Lambda + h^{\vee}\Lambda_0)$
of $\hhg$ and the defining relations of
$L(\Lambda + h^{\vee}\Lambda_0)$ are presented
in detail in the Appendix.
With the help of the theory developed in
Sec.1, we prove in Subsec.2.2 that the SVOA $L_{k,0}$ is
rational for positive integral $k$ and that the
representations and fusion rules
for the SVOA $L_{k,0}$ are in 1-1 correpondence with those for the VOA
$\bar{L}_{k,0}$.

In Subsec.3.1 we construct $N=1$ SVOAs $M_c$
and $V_c$ corresponding to the representations of
the Neveu-Schwarz algebra. We then discuss the rationality
and the fusion rules of $V_c$.

In Sec.4 we construct the SVOAs generated by
charged and neutral free fermionic fields. We prove
that such an SVOA
is rational and has a unique
irreducible representation, namely itself.

{\bf Acknowledgement.} We thank Shun-Jen Cheng for useful
discussions.

\section {General constructions and theorems}
\setcounter{equation}{0}

\subsection {\em Definitions}
For a rational function $f(z,w)$, with possible poles only at
$z = w, z = 0$ and $ w = 0$, we denote by
$\iota_{z, w}f(z,w)$  the power series expansion
of $f(z, w)$ in the domain $\mid z\mid >\mid w \mid$.
Set $\Bbb Z_{+} = \{ 0,1,2,\cdots\},\, \Bbb N = \{1,2,3,\cdots\}.$

A superalgebra is an algebra $V$ with a $\Bbb Z_2$-gradation
$V = V_{\bar{0}} \oplus V_{\bar{1}}$. Elements in $V_{\bar{0}}
\,\,({\em resp.}V_{\bar{1}})$ are called even ({\em resp.} odd).
Let $\tilde{a}$ be 0 if $a \in V_{\bar{0}}$,
and 1 if $a \in V_{\bar{1}}$.
The general principle to extend identities in
VOAs to SVOAs is the usual one:
if in certain formulas of VOAs there are some
monomials of vertex operators with interchanged terms,
then in the corresponding formulas in SVOAs every interchange
of neighboring terms, say $a$ and $b$, is accompanied by
multiplication of the monomial by the factor
$ (-1)^{\tilde{a} \tilde{b}} $.

\begin{definition}
A {\em vertex operator superalgebra} is a
$\frac{1}{2}\Bbb Z_{+}$-graded vector space $V=\bigoplus_
{n \in\frac{1}{2}\Bbb Z_{+}}V_n$ with
a sequence of linear operators $\{a(n)\mid n\in \Bbb
Z\}\subset End\ V$ associated to every $a\in V$,
whose generating series
$Y(a,z)=\sum_{n\in \Bbb Z}a(n)z^{-n-1}\in (End\ V)[[z,z^{-1}]]$,
called the {\em vertex operators} associated to $a$,
satisfy the following axioms:
\begin{enumerate}
\item[{\bf (A1)}]  $Y(a,z)=0$ iff $a=0$.

\item[{\bf (A2)}] There is a {\em vacuum} vector, which we denote by 1,
such that
$$Y(1,z)=I_V\,(I_V\, \mbox{is the identity of}\,\, End\,V). $$

\item[{\bf (A3)}]
There is a special element $\omega \in V$
(called the  {\em Virasoro\ element}), whose
 vertex operator we write in the form
$$Y(\omega,z)=\sum_{n\in \Bbb Z}\omega(n)z^{-n-1}=\sum_{n\in \Bbb
Z}L_nz^{-n-2},$$ such that
$$L_0\mid_{V_n}=nI\mid_{V_n},$$
\begin{equation}
Y(L_{-1}a,z)=\frac{d}{dz}Y(a,z)\ \ \mbox{for every } a \in V ,
\end{equation}
\begin{equation}
[L_m,L_n]=(m-n)L_{m+n}+\delta_{m+n,0}\frac{m^3-m}{12}c, \label{eq_vir}
\end{equation}
where $c$ is some constant in $\Bbb{C}$, which is called the {\em rank} of $V$.

\item[{\bf (A4)}] The {\em Jacobi identity} holds, i.e.
\begin{eqnarray*}
& & \mbox{Res}_{z-w} \bigl( Y(Y(a,z-w)b,w) \iota_{w,z-w} ((z-w)^m z^n)) \\
& = & \mbox{Res}_{z} \bigl( Y(a,z)Y(b,w)\iota_{z,w}(z-w)^{m}z^n )\\
 & & - (-1)^{\tilde{a}\tilde{b}}\mbox{Res}_{z}
\bigl( Y(b,w)Y(a,z)\iota_{w,z}(z-w)^{m}z^n)
\end{eqnarray*}
for any $m,n\in \Bbb Z$.
\end{enumerate}
\label{definition_svoa}
\end{definition}

An element $a\in V$ is called ${\em homogeneous}$ of degree $ n$
if $a$ is in $V_n$.
In this case we write $\deg a=n.$

Define a natural $\Bbb Z_2$-gradation of $V$
by letting
$$V_{\bar{0}} = \bigoplus_{n \in \Bbb Z_{+}}V_{n},\,\,\,\,\,
V_{\bar{1}} = \bigoplus_{n \in\frac{1}{2} + \Bbb Z_{+}}V_{n}.$$
$V = V_{\bar{0}} + V_{\bar{1}}.$
$V_{\bar{0}}\, ({\em resp.}\,\,\,V_{\bar{1}})$
is called the even\, ({\em resp.}\,\, odd) part of $V$.
Elements in $V_{\bar{0}} \,({\em resp.}\,\,V_{\bar{1}})$ are called
even\, ({\em resp.}\,\,odd).

We now introduce the notion of an $N=1$ SVOA.
\begin{definition}
$V$ is called an {\em $N=1$ (NS-type) SVOA} if
axiom (A3) is replaced by the following stronger axiom:
\begin{enumerate}
\item[${\bf (A3^{'})}$]
There is a special element $\tau \in V$
(called the  {\em Neveu-Schwarz element}), whose corresponding
 vertex operator we write in the form
$$Y(\tau,z) = \sum_{n\in \Bbb Z}\tau(n)z^{-n-1}=
\sum_{n\in \Bbb Z}G_{n+\frac{1}{2}}z^{-n-2},$$
such that the element $\omega := \frac{1}{2}G_{-\frac{1}{2}}\tau$
satisfies $(A3)$, and the
commutation relations
\begin{eqnarray*}
[G_{m+\frac{1}{2}},L_n] = (m+\frac{1}{2}-\frac{n}{2})G_{m+n+\frac{1}{2}},
\label{eq_ns1}
\end{eqnarray*}
\begin{eqnarray*}
[G_{m+\frac{1}{2}},G_{n-\frac{1}{2}}]_{+} =
2L_{m+n}+\frac{1}{3}m(m+1)\delta_{m+n,0}c,\,m,n \in \Bbb Z \label{eq_ns2}
\end{eqnarray*}
\end{enumerate}
also hold.
\label{definition_svoa1}
\end{definition}

We list some properties of $SVOA$s which are
anologous to those in the VOA case. For more detail see [FLM].

\begin{equation}
[a(n),Y(b,z)]_{\mp} = \sum_{i \geq 0}
{n \choose i} z^{n-i}Y(a(i)b,z),
\end{equation}
$$[L_0,Y(a,z)] = (z\frac{d}{dz}+\deg a)Y(a,z),$$
\begin{equation}
[L_{-1},Y(a,z)] = \frac{d}{dz}Y(a,z),
\end{equation}
\begin{equation}
a(n)V_m \subset V_{m+\deg a -n-1}, \label{eq_212}
\end{equation}
$$ Y(a,z)1 = e^{zL_{-1}}a,$$
$$ Y(a,z)b = (-1)^{\tilde{a}\tilde{b}}e^{zL_{-1}}Y(b,-z)a,$$
$$ a(n)1 = 0,\mbox{ for } n\geq 0,$$
$$ a(-n-1)1 = \frac{1}{n!}L_{-1}^n a\,\,\mbox{for}\,\, n \geq 0.$$
Moreover, $N=1$ SVOAs have the extra property that:
$$ [G_{-\hf},Y(a,z)]_{\mp}=Y(G_{-\hf}a,z).$$

\begin{definition}
Given an $SVOA$  $V$, {\em a representation of $V$
(or V-module)} is a $\frac{1}{2}\Bbb Z_{+} $-graded vector space
 $M=\bigoplus_{n \in\frac{1}{2}\Bbb Z_{+}}M_{n}$ and
a linear map $$V\longrightarrow (End\ M)[[z,z^{-1}]],$$
$$a\longmapsto Y_M (a,z)=\sum_{n\in \Bbb Z}a(n)z^{-n-1},$$
satisfying
\begin{enumerate}
\item[{\bf (R1)}]  $a(n)M_m \subset M_{m+\deg a -n-1}\,\,\, \mbox{for every
homogeneous element a.}$

\item [{\bf (R2)}]  $Y_M(1,z)=I_M,$ and setting
$Y_M(\omega,z)=\sum_{n\in \Bbb Z}L_nz^{-n-2},$ we have

$ [L_m,L_n]=(m-n)L_{m+n}+\delta_{m+n,0}\frac{m^3-m}{12}c,$

$Y_M(L_{-1}a,z)=\frac{d}{dz}Y_M(a,z)\,\,
\mbox{for every}\,\,\,a \in V.$

\item[{\bf (R3)}]  The {\em Jacobi identity} holds, i.e.
\begin{eqnarray*}
& & \mbox{Res}_{z-w} \bigl( Y_M (Y(a,z-w)b,w) \iota_{w,z-w}
((z-w)^m z^n))  \\
& = & \mbox{Res}_{z} \bigl( Y_M (a,z)Y_M (b,w)
\iota_{z,w}(z-w)^{m}z^n ) \\
&& - (-1)^{\tilde{a}\tilde{b}}
\mbox{Res}_{z} \bigl( Y_M (b,w)Y_M (a,z)\iota_{w,z}(z-w)^{m}z^n)
\end{eqnarray*}
for any $m,n\in \Bbb Z$.
\end{enumerate}
\label{definition_module}
\end{definition}

\begin{definition}
Given an $N=1\,\,SVOA$ $V$, $M$ is called
{\em a representation of $V$} if
axiom (R2) is replaced by the following stronger axiom:
\begin{enumerate}
\item[${\bf (R2')}$]
Set
$Y_M (\tau,z) =
\sum_{n\in \Bbb Z}G_{n+\frac{1}{2}}z^{-n-2}$ and
$\omega := \frac{1}{2}G_{-\frac{1}{2}}\tau$ .
Then $\omega$ satisfies (R2), and the
commutation relations
$$[G_{m+\frac{1}{2}},L_n] = (m+\frac{1}{2}-\frac{n}{2})G_{m+n+\frac{1}{2}},$$
$$[G_{m+\frac{1}{2}},G_{n-\frac{1}{2}}]_{+} =
2L_{m+n}+\frac{1}{3}m(m+1)\delta_{m+n,0}c,\,m,n \in \Bbb Z$$
\end{enumerate}
also hold.
\label{definition_module1}
\end{definition}

The notions of submodules, quotient modules, submodules generated
 by a subset, direct sums, irreducible modules,
 completely reducible modules, etc., can
be introduced in the usual way. As
a module over itself, $V$ is called the {\em adjoint module}. A submodule
of the adjoint module is called an {\em ideal} of $V$.
Given an ideal $I$ in $V$ such that $1 \not \in I, \omega \not \in I,$
the quotient $V/I$ admits a natural SVOA structure.

\begin{definition}
A $SVOA$ is called {\em rational}
if it has finitely many irreducible modules
and every module is
 a direct sum of irreducibles.
\label{definition_rational}
\end{definition}
We will now extend the definition of intertwining
operators and fusion rules of representations of
VOAs ( [FHL] ) to SVOAs.

For simplicity, we will only define an intertwining operator for
$V$-modules $M^i = \oplus_{n \in \hf \Bbb Z_{+}}M^i(n),
\,(i = 1,2,3)$,
satisfying $L_0 \mid_ {M^i (n)} = (h_i + n)I\mid_ {M^i (n)},$ for some
complex numbers $h_1, h_2, h_3.$ We define a
$\Bbb Z_2$-gradation of $M^i$ by letting $\tilde{v} = 0$
if $ v \in M^i (n), n \in \Bbb Z$;  $\tilde{v} = 1$
if $ v \in M^i (n), n \in \hf + \Bbb Z$.

\begin{definition}
Under the above assumptions, an
intertwining operator of type
$ { {\,\,\,\,\,\,\,\,M^3 \,\,\,\,\,\,\,\,} \choose
{M^1 \,\,\,\,\,\,\,\, M^2}}  $
is a linear map
$$ I(\cdot,z) : v \mapsto \sum_{k \in I}
v(n) z^{-n-1+(h_3 -h_1 -h_2)},\,\,v \in M^1,\,\,
v(n) \in Hom_{\Bbb C}(M^2,M^3) $$
satisfying
\begin{enumerate}
\item[{\bf (I1)}]  For homogeneous $v \in M^1,$
$$I (L_{-1}v,z)=\frac{d}{dz}I(v,z)\ \ \mbox{for every } v\in M^1 , $$
\item[{\bf (I2)}]
For any $a \in V, v \in M^1,$ and $m,n\in \Bbb Z$,
\begin{eqnarray*}
& & \mbox{Res}_{z-w} \bigl( I (Y(a,z-w)v,w) \iota_{w,z-w}
((z-w)^m z^n))  \\
& = & \mbox{Res}_{z} \bigl( Y(a,z)I (v,w)
\iota_{z,w}(z-w)^{m}z^n )\\
 & & - (-1)^{\tilde{a}\tilde{v}}
\mbox{Res}_{z} \bigl( I(v,w)Y (a,z)\iota_{w,z}(z-w)^{m}z^n).
\end{eqnarray*}
\end{enumerate}
\label{definition_intertwin}
\end{definition}

We denote by $ I { {\,\,\,\,\,\,\,\,M^3 \,\,\,\,\,\,\,\,} \choose
{M^1 \,\,\,\,\,\,\,\, M^2}}$
the vector space of intertwining operators of type
${ {\,\,\,\,\,\,\,\,M^3 \,\,\,\,\,\,\,\,} \choose
{M^1 \,\,\,\,\,\,\,\, M^2}} .$

An immediate consequence of this definition is that
for homogeneous $v \in M^1,$
$$ v(n) M^2_m \subset M_{m +\deg v -n-1}^3,$$
where $\deg v = k $ means that $ v \in M_k^1.$

We now assume that $V$ is a rational SVOA and $\{M^i,\,\,i \in J \}$
is the complete set of the irreducible modules of $V$.
Denote by $N_{\,k}^{ij}$ the dimension of the vector
space $ I { {\,\,\,\,\,\,\,\,M^k \,\,\,\,\,\,\,\,} \choose
{M^i \,\,\,\,\,\,\,\, M^j}}$. We define the fusion rules
as the formal product rules
$$M^i \times M^j = \sum_{k \in J} N_{\,k}^{ij} M^k.$$

\subsection{\em The associative algebra A(V) and related theorems}

\begin{definition}
We define bilinear maps $*: V\times V\rightarrow V$,
$\circ : V\times V\rightarrow V$as follows. For homogeneous $a,b$, let
$$ a*b = \left\{ \begin{array}{ll}
{Res_z \left(Y(a,z)\frac{(z+1)^{\deg a}}{z} b\right)},&{\mbox{if}
\,\,\,a,b \in V_{\bar{0}},}
\\0,&{\mbox{if}\,\,\, a \,\,\,\mbox{or}\,\,\,
b \in V_{\bar{1}}.}\end{array} \right.$$
$$ a \circ b = \left\{ \begin{array}{ll}
{Res_z \left(Y(a,z)\frac {(z+1)^{\deg a}}{z^2}b\right)},&
{\mbox{for}\,\,\, a \in V_{\bar{0}}} \\
Res_z\left(Y(a,z)\frac {(z+1)^{\deg a-\hf}}{z}b\right),&
{\mbox{for}\,\,\, a \in V_{\bar{1}}.}
\end{array} \right.$$
Extend to $V\times V$ by linearity,
denote by $O(V)\subset V$ the linear span of elements
of the form $ a \circ b $, and by $A(V)$ the quotient space $V/O(V)$.
\label{definition_star}
\end{definition}

\begin{remark}
\begin{enumerate}
\item[1)]
$O(V)$ is a $\Bbb Z_2$-graded subspace of $V$.
\item[2)]
If $a \in V_{\bar{1}},$ then
$$a \circ 1 = Res_z \left(Y(a,z)\frac {(z+1)^{\deg a-\hf}}{z}1\right)=a.$$
Hence $O(V) =O_{\bar{0}}(V) + V_{\bar{1}}$, where
$O_{\bar{0}}(V) = O_(V)\cap V_{\bar{0}}$.
Thus $A(V) = V_{\bar{0}}/O_{\bar{0}}(V)$.
Denote by  $O_e (V)\,({\em resp.}\,O_d (V))$ the linear span of
the elements\, $a \circ b$\, for $a,b \in V_{\bar{0}}
\,({\em resp.}\, V_{\bar{1}}).$ The intersection
$O_e (V) \cap O_d (V)$ need not be empty.
\label{remark_noodd}
\end{enumerate}
\end{remark}

It is convenient to introduce an equivalence relation
$\sim$ as follows.
 For $a, b \in V, a \sim b$ means
$a - b \equiv 0 \,mod\, O(V)$. For $f, g \in End\ V, f \sim g$
 means $f \cdot c \sim g \cdot c $ for any $c \in V$.
Let $[a]$ to denote the image of $a$ in $V$ under the
projection of $V$ onto $A(V)$.

\begin{lemma}
\begin{enumerate}
\item[1)]  $L_{-1}a + L_0 a\sim 0$ if $a\in V_{\bar{0}}$.

\item[2)]  For every homogeneous element $a\in V$, and
$m\geq n\geq 0$, one has
$$Res_z \left((Y(a,z)\frac {(z+1)^{\deg a+n}}{z^{2+m}}\right)\sim 0,
\mbox{if}\,\,\, a \in V_{\bar{0}}.$$
$$Res_z \left((Y(a,z)\frac {(z+1)^{\deg a+n-\hf}}{z^{1+m}}\right)\sim 0,
\mbox{if}\,\,\, a \in V_{\bar{1}}.$$

\item[3)]  For any homogeneous element $a,b \in V_{\bar{0}}$, one has
$$ a*b  \sim Res_z \left(Y(b,z)
\frac {(z+1)^{\deg b-1}}{z}a \right). $$
\end{enumerate}
\label{lemma_keylemma}
\end{lemma}

{\em Proof.}  Noting that $V_{\bar{0}}$ is a vertex operator algebra,
we see that 1), the first part of 2) and 3) are
the same as Lemma 2.1.1, 2.1.2 and 2.1.3 in [Z]. The proof of the
second part of 2) is similar to that of the first part. $\Box$

The following theorem is an analog of Theorem 2.1.1 in [Z].

\begin{theorem}
\begin{enumerate}
\item[1)]  $O(V)$ is a two-sided ideal of $V$ under
the multiplication $*$. Moreover, the quotient algebra
$(A(V),*)$ is associative.
\item[2)]  [1] is the unit element of the algebra $A(V)$.
\item[3)]  $[\omega]$ is in the center of $A(V)$.
\item[4)]  $A(V)$ has a filtration $A_0 (V)\subset A_1 (V)
\subset \cdots$, where $A_n(V)$ is the image of
$\oplus_{i \in \hf \Bbb Z_{+},i \leq n} V_i.$
\end{enumerate}
\label{theorem_associativity}
\end{theorem}

{ \em Sketch of a proof.} To prove 1), it is enough to prove the
following relations:
$$O_{\bar{0}}(V)*V \subset O(V), $$
$$ V_{\bar{0}}*O_{\bar{0}}(V) \subset O(V), $$
$$ (a*b)*c-a*(b*c) \in O(V). $$
By the definition of the operation $*$ and Remark \ref{remark_noodd},
it suffices to prove that for homogeneous $a,b,c$ one has
\begin{equation}
(a \circ b) * c \subset O(V)\, \mbox{for} \, a,b,c \in V_{\bar{0}},
\label{eq_1}
\end{equation}
\begin{equation}
a * (b \circ c)\subset O(V)\,
\mbox{for} \, a,b,c \in V_{\bar{0}},\label{eq_2}
\end{equation}
\begin{equation}
(a \circ b) * c \subset O(V)\,
\mbox{for} \, a,b \in V_{\bar{1}},\label{eq_3}
\end{equation}
\begin{equation}
a * (b \circ c)\subset O(V)\,
\mbox{for} \, a\in V_{\bar{0}}, b,c \in V_{\bar{1}},
\label{eq_4}
\end{equation}
\begin{equation}
(a*b)*c-a*(b*c) \in O(V)\,
\mbox{for} \,a,b,c \in V_{\bar{0}}.\label{eq_5}
\end{equation}
The proofs of  (\ref{eq_1}), (\ref{eq_2}) and
(\ref{eq_5}) are the same as in the VOA cases (see the proof
of Theorem 2.1.1 in [Z]).

To prove (\ref{eq_3}), for $a,b \in V_{\bar{1}},c \in V$ homogeneous,
we have
\begin{eqnarray*}
\lefteqn{(a \circ b) * c} \\
&=&\strhf *c \\
&=&\sum_{i=0}^{\deg a-\hf} \binomhf \left(a(i-1)b \right)*c \\
&=&\sum_{i=0}^{\deg a-\hf} \binomhf Res_w
\left(Y(a(i-1)b,w)\frac {(w+1)^{\deg a+\deg b-i}} {w}c\right) \\
&=&\sum_{i=0}^{\deg a-\hf} \binomhf \times \\
 && \hspace{0.8 cm} \times Res_w Res_{z-w}
\left(Y(Y(a,z-w)b,w)(z-w)^{i-1}\frac {(w+1)^{\deg a+\deg b-i}} {w}c\right) \\
&=&Res_w Res_{z-w}\left(Y(Y(a,z-w)b,w)\frac {(z+1)^{\deg
a-\hf}(w+1)^{\deg b+\hf }} {w(z-w)}c\right) \\
&=&Res_z Res_w\left(Y(a,z)Y(b,w)\frac {(z+1)^{\deg
a-\hf}(w+1)^{\deg b +\hf}} {w(z-w)}c\right) \\
&&+Res_w Res_z \left(Y(b,w)Y(a,z)\frac {(z+1)^{\deg
a-\hf}(w+1)^{\deg b+\hf }} {w(z-w)}c\right) \\
&=&\sum_{i\in \Bbb Z_{+}} Res_z Res_w \left(Y(a,z)Y(b,w)
z^{-1-i}w^i \frac {(z+1)^{\deg a-\hf}(w+1)^{\deg b+\hf }} {w}c\right) \\
&&-\sum_{i\in \Bbb Z_{+}} Res_w Res_z \left(Y(b,w)Y(a,z)
w^{-1-i}z^i \frac {(z+1)^{\deg a-\hf}(w+1)^{\deg b+\hf }} {w}c\right).
\end{eqnarray*}

By Lemma \ref{lemma_keylemma} the right hand side
of the last identity is in $O(V)$.

To prove (\ref{eq_4}),
for $a\in V_{\bar{0}}, b,c \in V_{\bar{1}}$ homogeneous, we have

\begin{eqnarray*}
\lefteqn{a * (b \circ c) - b \circ (a * c)} \\
&=&  Res_z \left(Y(a,z)\frac{(z+1)^{\deg a}}{z}\right)
            Res_w \left(Y(b,w)\frac{(w+1)^{\deg b-\hf}}{w}c\right)  \\
&&- Res_w \left(Y(b,w)\frac{(w+1)^{\deg b-\hf}}{w}\right)
                 Res_z \left(Y(a,z)\frac{(z+1)^{\deg a}}{z}c\right) \\
&=& Res_w Res_{z-w} \left(Y(Y(a,z-w)b,w)\frac{(z+1)^{\deg a}}{z}
                \frac{(w+1)^{\deg b-\hf}}{w}c\right)  \\
&=& \sum_{i=0}^{\deg a}\sum_{j\in \Bbb Z_{+}}\binom
Res_w\left(Y(a(i+j)b,w)(-1)^j \frac{(w+1)^{\deg a+\deg
b-i-\hf}}{w^{j+2}}c \right).
\end{eqnarray*}

Since $\deg (a(i+j)b)=\deg a+ \deg b-i-j-1$, and $a(i+j)b\in V_{\bar{1}}$,
by Lemma \ref{lemma_keylemma}, the right-hand side of the last identity is
in $O(V)$. The second term of the
left-hand side is also in $O(V)$ by definition. Then so is the first term.

The proof of statements 2), 3) and 4) is the same as in the VOA case.
(For details see the proof of Theorem 2.1.1 in [Z]).  \,\,\, $\Box$

The following proposition follows from the definition of
$A(V)$.
\begin{proposition}
Let $I$ be an ideal of $V$ with the
$\Bbb Z_{2}$-gradation $I_{\bar{0}} \oplus I_{\bar{1}}$
consistent with that of $V$.
Assume $1 \not \in I,
\omega \not \in I$. Then the associative algebra
$ A(V/I)$ is isomorphic to $A(V) / {[I]}$, where $[I]$ is the image
 of $I$ in $A(V)$.
\label{proposition_associativity}
\end{proposition}

For any homogeneous
$a \in V_{\bar{0}}$ we define $o(a) = a(\deg a-1)$
and extend this map linearly to $V_{\bar{0}}$.
It follows from (\ref{eq_212}) that $ o(a)M_n \subset M_n$.
In particular, $o(a)$ maps $M_0 $ into itself. We may assume
that $M_0 \neq 0$ without loss of generality.

\begin{theorem}
Let $\M$ be a V-module. Then $M_0$ is an $A(V)$-module defined
as follows: for $[a] \in A(V)$, let $a\in V_{\bar{0}}$ be a
preimage of $[a]$. Then $[a]$ acts on $M_0$ as $o(a)$.
\label{theorem_toplevel}
\end{theorem}

{\em Proof.} An equivalent way to state this theorem is that
for $a,b \in V_{\bar{0}}$,
$o(a)o(b)\mid_{M_0}=o(a*b)\mid_{M_0}$, and for
$c\in O(V)=O_d (V)+O_e (V)$,
$o(c)\mid_{M_0}=0$. We only need to prove that
$o(c)\mid_{M_0}=0$ for $c\in O_d (V)$ since $V_{\bar{0}}$ is
a vertex operator algebra and so the rest of the
statements above holds (For details see Theorem 2.1.2 and its
proof in [Z]).

Given $a,b \in V_{\bar{1}}$ homogeneous, we have
\begin{eqnarray*}
\lefteqn{o(a \circ b)} \\
&=& o(\strhf) \\
&=& \sum_{i=0}^{\deg a-\hf}\binomhf o\left(a(i-1)b\right) \\
&=& \sum_{i=0}^{\deg a-\hf}\binomhf \left(a(i-1)b\right)(\deg a+\deg b-i-1)\\
&=& Res_w Res_{z-w}\sum_{i=0}^{\deg a-\hf}\binomhf \times \\
 &&\hspace{2 cm}\times \left(Y((a,z-w)b,w)
(z-w)^{i-1}w^{\deg a+\deg b-i-1}\right)  \\
&=& Res_w Res_{z-w}\left(Y((a,z-w)b,w)\frac{z^{\deg a-\hf}
w^{\deg b-\hf}}{z-w}\right)  \\
&=& Res_z Res_w \left(Y(a,z)Y(b,w)\frac{z^{\deg a-\hf}w^{\deg
b-\hf}}{z-w}\right)  \\
&&+ Res_w Res_z \left(Y(b,w)Y(a,z)\frac{z^{\deg a-\hf}
w^{\deg b-\hf}}{z-w}\right)  \\
&=& \sum_{i\in \Bbb Z_{+}} Res_z Res_w \left(Y(a,z)Y(b,w)z^{\deg a-i-\thf}
w^{\deg b-\hf+i}\right)   \\
&&- \sum_{i\in \Bbb Z_{+}} Res_w Res_z \left(Y(b,w)Y(a,z)z^{\deg a+i-\hf}
w^{\deg b-i-\thf}\right) \\
&=& \sum_{i\in \Bbb Z_{+}} a(\deg a-i-\thf)b(\deg b+i-\hf)\\
&& -\sum_{i\in \Bbb Z_{+}} b(\deg b-i-\thf)a(\deg a+i-\hf).
\end{eqnarray*}

The right-hand side of the above identities acting on $M_0$ is 0
since
$$a(\deg a+i-\hf)\mid_{M_0}=b(\deg b+i-\hf)\mid_{M_0}=0. \,\, \,\,\Box $$

\begin{theorem}
Given an $A(V)$-module
$(W,\pi)$, there exists a V-module $\M$ such that the
$A(V)$-modules $M_0$ and $W$ are isomorphic.
Moreover, this gives a bijective
correspondence between the set of irreducible $A(V)$-modules
and the set of irreducible V-modules.
\label{theorem_correspondence}
\end{theorem}

{\em Sketch of a proof.} First we have the following recurrent
formula for $n$-correlation functions on $\langle M_0,(M_0)^* \rangle$
for a given $V$-module $M=\oplus_{i \in \hf \Bbb Z_{+}} M_i$, where
$M_0^*$ is the dual space of $M_0$.
(The proof is similar to that of Lemma 2.2.1 in [Z].)
Given $v \in M_0,\,\, v' \in M_0^*$, and
homogeneous $a_1 \in V$, we have
\begin{eqnarray*}
\lefteqn{\langle v',\,\,Y(a_1,z_1)Y(a_2,z_2)\cdots Y(a_m,z_m)v \rangle}\\
&=& \left\{\begin{array}{l}
\sum_{k=2}^m \sum_{i \in \Bbb Z_{+}}
(-1)^{(\tilde{a}_2+\cdots+\tilde{a}_{k-1})}
F_{\deg a_1 -\hf,i}(z_1,z_k)\times \\
\,\,
\times \langle v',Y(a_2,z_2)\cdots
Y(a_1(i)a_k,z_k)\cdots Y(a_m,z_m)v \rangle \,\,
\mbox{if}\,\,a_1 \in V_{\bar{1}}, \\
\sum_{k=2}^m \sum_{i \in \Bbb Z_{+}}
F_{\deg a_1,i}(z_1,z_k)\times \\
\,\,\,\,\,\,
\times \langle v',Y(a_2,z_2)\cdots
Y(a_1(i)a_k,z_k)\cdots Y(a_m,z_m)v \rangle  \\
\,\,
+ z_1^{-\deg a_1}\langle a_1(\deg a_1-1)^* v',
Y(a_2,z_2)\cdots Y(a_m,z_m)v \rangle \,\,
\mbox{if}\,\,a_1 \in V_{\bar{0}},
\end{array}\right.
\end{eqnarray*}
where $F_{\deg a,i}$ is defined by
\begin{eqnarray*}
F_{n,i}(z,w) &=&
\sum_{j\in \Bbb Z_{+}} {{n+j} \choose {i}} z^{-n-j}w^{n+j-i} \\
&=& \iota_{z,w}\left(z^{-n}\frac {1}{i!}
(\frac{d^i}{dw^i})\frac{w^{n}}{z-w}\right).
\end{eqnarray*}
This recurrent formula means that the $n$-correlation functions
on $\langle M_0,(M_0)^* \rangle$ are determined by the $A(V)$-module
structure on $M_0$. The completion of the proof of this theorem
is similar to that in Theorem 2.2.1 in [Z].   $\Box$

\begin{remark}
Thus we have a functor from the category of $V$-modules
to the category of $A(V)$-modules which is bijective
on the sets of irreducibles.
\end{remark}
\subsection{\em Fusion rules}

In this subsection, to generalize the construction
of [FZ], we define a bimodule $A(M)$ of $A(V)$ for every
$V$-module $M$. We then give a description of the
fusion rules in terms of $A(M)$.
The proofs are only sketched.

\begin{definition}
For a $V$-module $M$, we define bilinear operations
$a*v$ and $v*a$, for $a\in V$ homogeneous and $ v \in M$, as follows
\begin{equation}
a*v = Res_z \left(Y(a,z)\frac{(z+1)^{\deg a}}{z} v\right),\,
\mbox{for}\,\,\, a\in V_{\bar{0}}, \label{eq_left}
\end{equation}
\begin{equation}
v*a = Res_z \left(Y(a,z)\frac{(z+1)^{\deg a-1}}{z} v\right),\,
\mbox{for}\,\,\, a\in V_{\bar{0}}, \label{eq_right}
\end{equation}
$$a*v = 0,\,\,v*a = 0,\,\,\,\mbox{for}\,\,a\in V_{\bar{1}},$$
and extend linearly to $V$.
We also define $O(M)\subset M$ to be the linear span of elements
of the forms
$$Res_z \left(Y(a,z)\frac{(z+1)^{\deg a}}{z^2} v\right),\,\,\,
\mbox{for}\,\,\, a\in V_{\bar{0}} \,\,\mbox{and}$$
$$Res_z \left(Y(a,z)\frac{(z+1)^{\deg a-\hf}}{z} v\right),\,\,\,
\mbox{for}\,\,\, a\in V_{\bar{1}}.$$
Let $A(M)$ be the quotient space $M/O(M)$.
\label{definition_A(M)}
\end{definition}

We have the following theorem which is an analogue of Theorem 1.5.1
in [Z].

\begin{theorem}
$A(M)$ is an $A(V)$-bimodule with the left action of
$A(V)$ defined by (\ref{eq_left}) and the right action
by (\ref{eq_right}). Moreover the left and right action
of $A(V)$ commute with each other.
\end{theorem}

{\em Sketch of a proof.} By a similar
argument to Lemma \ref{lemma_keylemma}, we see that
$$Res_z (Y(a,z)\frac{(z+1)^{\deg a+n}}{z^{2+m}} v)\in O(M), \,
\mbox{for}\, a\in V_{\bar{0}},$$
$$Res_z (Y(a,z)\frac{(z+1)^{\deg a+n-\hf}}{z^{1+m}} v)\in O(M),\,
\mbox{for}\, a\in V_{\bar{1}},$$
for $m \geq n \geq 0$, $v \in M$.

Recall that $O(V) = O_d (V) + O_e (V)$. To prove the theorem,
we need to check that
\begin{equation}
O_d (V) * v \subset O(M),\,\,v * O_d (V)  \subset O(M), \label{eq_6}
\end{equation}
\begin{equation}
O_e (V) * v \subset O(M),\,\,v * O_e (V)  \subset O(M), \label{eq_7}
\end{equation}
\begin{equation}
a* O(M) \subset O(M),\,\,O(M) * a \subset O(M),\label{eq_8}
\end{equation}
\begin{equation}
(a*b) * v - a * (b * v) \in O(M), \label{eq_9}
\end{equation}
\begin{equation}
(v * a) * b -v * (a *b) \in O(M), \label{eq_10}
\end{equation}
\begin{equation}
(a * v) * b - a * (v * b) \in O(M). \label{eq_11}
\end{equation}

The proof of (\ref{eq_6}) is
similar to that of Theorem \ref{theorem_associativity}.
The proofs of (\ref{eq_7}), (\ref{eq_8}), (\ref{eq_9}), (\ref{eq_10})
and (\ref{eq_11}) are similar to those in [Z]. \,\,\,\,   $\Box$

Consider left $V$-modules
$M^i= \oplus_{n \in \hf \Bbb Z_{+}} M^i(n),\,\,\,
i=1,2,3$. Note that $M^2 (0)$ is a left module over $A(V)$,
$\left(M^3 (0)\right)^{*}$ is
a right module over $A(V)$, and $A(M^1)$ is a bimodule
over $A(V)$. Hence we can consider the tensor product
$ M^3 (0)^{*} \otimes_{A(V)} A(M^1) \otimes_{A(V)} M^2 (0) $
of $A(V)$-modules.

The following theorem is an analogue of Theorems 1.5.2 and 1.5.3
in [FZ].

\begin{theorem}
Let $M^i = \sum_{n\in \hf \Bbb Z_{+}}M^i (n)(i=1,2,3)$
be $V$-modules, satisfying
$L_0 \mid_ {M^i (n)} = (h_i + n)I \mid_ {M^i (n)},$ for some
complex numbers $h_1, h_2, h_3.$
\begin{enumerate}
\item[1)]  Let $I(\cdot,z)$ be an intertwining
operator of type ${ {\,\,\,\,\,\,\,\,M^3 \,\,\,\,\,\,\,\,} \choose
{M^1 \,\,\,\,\,\,\,\, M^2}}.$ Then
$\langle v'_3, o(v_1)v_2 \rangle $ defines a linear functional $f_{I}$
on $ M^3 (0)^{*} \otimes_{A(V)} A(M^1) \otimes_{A(V)} M^2 (0), $
where $v'_3 \in M^3 (0)^{*}$, $v_1 \in M^1, v_2 \in M^2$,

\item[2)]  The map $I \mapsto f_I$ given in 1) defines an
isomorphism of vector spaces
$I { {\,\,\,\,\,\,\,\,M^3 \,\,\,\,\,\,\,\,} \choose
{M^1 \,\,\,\,\,\,\,\, M^2}}$ and
$ (M^3 (0)^{*} \otimes_{A(V)} A(M^1) \otimes_{A(V)} M^2 (0))^{*} $
if
$M^i \,(i = 1,2,3) $ are irreducible.
\end{enumerate}
\label{theorem_fusionrules}
\end{theorem}
{\em Proof.} The argument is similar to that in Theorem
\ref{theorem_correspondence}.   $\Box$

As a consequence, we obtain the following proposition,
which is an anologue of
Proposition 1.5.4 in [FZ].
\begin{proposition}
\begin{enumerate}
\item[1)]  Given a $V$-module $M$ and a submodule
$M^1$ of $M$, then the image $A(M^1)$ of $M^1$ in $A(M)$,
is a submodule of $A(V)$-bimodule $A(M)$, and the quotient
$A(M)/A(M^1)$ is isomorphic to the bimodule $A(M/M^1)$
corresponding to the quotient $V$-module $M/M^1$.

\item[2)]  If $I$ is an ideal of $V$, $1\not \in I,
\omega \not \in I$, and $I\cdot M \subset M^1$,
then $A(V/I)$-bimodule $A(M)/A(M^1)$ is isomorophic to
the $A(M/M^1)$.
\end{enumerate}
\label{proposition_fusionrules}
\end{proposition}

\begin{remark}
One can also consider the pre-SVOA (i.e. the SVOA which
may not admit a Virasoro element).
Similarly to the VOA case , one can still define the
associative algebra $A(V)$ and the $A(V)$-module
$A(M)$ for any $V$-module $M$ [L]. Theorems \ref{theorem_correspondence}
and \ref{theorem_fusionrules} are valid for the pre-SVOAs.
\end{remark}

\section {SVOA associated to representations
of affine Kac-Moody superalgebras}
\setcounter{equation}{0}

\subsection{\em $SVOA$ structures on $M_{k,0}$ and $L_{k,0}$}

In this subsection, we construct the SVOAs
associated to representations of affine Kac-Moody superalgebras
which are analogous to the construction of VOAs
associated to representations of affine algebras [FZ].
First let us recall some basic notions of  affine
Kac-Moody (super)algebras. Given a simple finite-dimensional
Lie algebra $\frak g$ of rank $l$ over $\Bbb C$,
we fix a Cartan subalgebra $\frak h$, a root
system $\Delta \subset {\frak {h}} ^{*} $ and a set of positive roots
$\Delta _{+} \subset \Delta$. Let $\frak g = \frak h
\,\,\bigoplus \,( \bigoplus_{\alpha \in
 \Delta} \frak g_{\alpha})$ be the root space decomposition
of $\frak g$. Let $e_i,\,f_i,\,h_i\,(i=1,\cdots,l)$
be the corresponding Chevalley generators. Denote by $\theta$ the
highest root and normalize the Killing form
$$ (\,,\,): \frak g \times \frak g \rightarrow \Bbb C $$
by the condition $(\theta ,\theta ) = 2 $. Let $\sigma$ be the
antilinear anti-involution
of $\frak g$. We choose $f_{\theta} \in \frak g _{-\theta}$ so that
$(f_{\theta},\sigma(f_{\theta})) = 1$, and set $e_{\theta} =
\sigma(f_{\theta})$.
We denote by $r_{\alpha}$ the
reflection with respect to $\alpha \in \Delta$
in the Weyl group $ W \in GL(\frak h) $ of $\frak g$..

The affine Kac-Moody superalgebra (of NS type)
is then defined by
$$ \hat{\hat{\frak g}} = \frak g \bigotimes \Bbb {C}[t,t^{-1},\xi]
\bigoplus \Bbb {C}{\bf k} \bigoplus \Bbb {C}d $$
with the following commutation relations
\begin{equation}
  [a(m), b(n)] = [a,b](m+n) + m\delta_{m+n, 0}(a,b){\bf k}, \label{eq_affine}
\end{equation}
\begin{equation}
[\bar{a}(m), \bar{b}(n)]_{+} = \delta_{m+n+1,0}(a,b){\bf k},
 \label{eq_clifford}
\end{equation}
\begin{equation}
[a(m), \bar{b}(n)] = \overline{[a,b]}(m+n),
\end{equation}
\begin{equation}
[{\bf k}, a(m)] = 0,\label{eq_affine1}
\end{equation}
\begin{equation}
[d, a(m)] = ma(m),
\end{equation}
\begin{equation}
[d, \bar{a}(m)] = (m+ \hf)\bar{a}(m),
\end{equation}
where $a,b \in \frak g, m,n \in \Bbb Z$,
$a(m) := a \otimes t^m,$
$\bar{a}(m) := a \otimes \xi t^m.$

Let
$$\bar{\frak g} = \frak g \bigotimes \xi$$
$$ \hat{\hat{\frak g}}_{+} = \frak g \bigotimes t\Bbb C [t] \bigoplus
\bar{\frak g}\bigotimes \Bbb C [t]$$
$$ \hat{\hat{\frak g}}_{-} = \frak g \bigotimes t^{-1}\Bbb C [t^{-1}]
\bigoplus
\bar{\frak g}\bigotimes t^{-1}\Bbb C [t^{-1}].$$
Then $\hat{\hat{\frak g}}_{+}$ and $ \hat{\hat{\frak g}}_{-}$
are subalgebras of $ \hat{\hat{\frak g}}$ and
$$\hat{\hat{\frak g}} = \hat{\hat{\frak g}}_{+} \bigoplus
\hat{\hat{\frak g}}_{-} \bigoplus \frak g
\bigoplus \Bbb C {\bf k} \bigoplus \Bbb {C}d .$$
Here we identify $\frak g \otimes 1$ with $\frak g $.
We let
$$ \hat{\frak h} = \frak h \bigoplus \Bbb C {\bf k}
\bigoplus \Bbb {C}d ,$$ and
extend the Killing form on $\frak h$ to $\hat{\frak h}$
by letting
$({\bf k}, d) =1, ({\bf k}, {\bf k}) = 0, (d,d) = 0,
(\Bbb C {\bf k} +\Bbb {C}d, \frak h) = 0.$
We identify $ \hat{\frak h}^{*} $ with
$ \hat{\frak h} $ using this bilinear form on $\hat{\frak h}$.

Given a $\frak g$-module V and a complex number $k$,
we can define the induced module $\tilde{V}_k$
over $\hhg$ as follows: $V$ can be viewed as a module over
$\hat{\hat{\frak g}}_{+} + \frak g + \Bbb C {\bf k} + \Bbb C d $
by letting $(\hat{\hat{\frak g}}_{+} \bigoplus \Bbb C d) V = 0$ and
${\bf k} =( k + h^{\vee})\, I \mid_V $. Then we let
$$\tilde{V}_k = \frak U (\hat{\hat{\frak g}})\bigotimes_{\frak U
(\hat{\hat{\frak g}}_{+} + \frak g + \Bbb C {\bf k} + \Bbb C d)}V.$$
Here and further $\frak U(\frak A)$ denotes the universal
enveloping algebra of a Lie (super)algebra $\frak A$.
In particular for any $\lambda \in \frak{h}^{*}$,
we let $L(\lambda)$ be the irreducible highest weight $\frak g$-module
with highest weight $\lambda$, and denote the $\hhg$-module
$\tilde{L}(\lambda)_k$
by $M_{k,\lambda}$. Let $J_{k,\lambda}$ be the maximal proper
submodule of the $\hat{\hat{\frak g}}$-module $M_{k,\lambda}.$
Denote $M_{k,\lambda}/J_{k,\lambda}$ by $L_{k,\lambda}$.
Note that if $\lambda = 0$, $L(0)$ is the trivial
$\frak g$-module $\Bbb C$ and $M_{k,0} \cong \frak U (\hat{\hat{\frak g}}_{-})$
as $\hat{\hat{\frak g}}_{-}$-modules.

Define a $\hf \Bbb Z$-gradation of $ \hat{\hat{\frak g}}$ by
the eigenvalues of $-d$:
$$\deg {\bf k} = 0,\,\,\, \deg a(n) = -n,\,\,\, \deg \bar{a}(n) = -n-\hf,
\,a \in \frak g.$$
This induces ${ \hf \Bbb{Z}}$-gradations of $\frak U (\hat{\hat{\frak g}}),
\frak U (\hat{\hat{\frak g}}_{-})$ and
${ \hf \Bbb Z_{+}}$-gradations of $M_{k,\lambda}$
if we let the degree of the highest weight of $L(\lambda )$
to be zero.
We denote the gradation decompositions by
$$\frak U (\hat{\hat{\frak g}}) = \bigoplus_{n\in \hf\Bbb Z}
\frak U (\hat{\hat{\frak g}})(n),
\frak U (\hat{\hat{\frak g}}_{-}) = \bigoplus_{n\in \hf\Bbb Z}
\frak U (\hat{\hat{\frak g}}_{-})(n),$$
$$M_{k,\lambda} = \bigoplus_{n\in \hf\Bbb Z_{+}}M_{k,\lambda}(n),$$
where $M_{k,\lambda} (n) = \frak U (\hat{\hat{\frak g}}_{-})(n)L(\lambda).$

Define a topological completion $\hat{\frak U} (\hat{\hat{\frak g}})$ of
$\frak U (\hat{\hat{\frak g}})$ as follows. Let
$$\frak U (\hat{\hat{\frak g}})_n^m = \sum_{i\leq m,i\in \hf \Bbb Z}
\frak U (\hat{\hat{\frak g}})_{n-i}\frak U (\hat{\hat{\frak g}})_i,
\,\,\,\,\mbox{for} \,m\in \hf \Bbb Z.$$
It is easy to see that
$$\frak U (\hat{\hat{\frak g}})_n^{m+\hf}\subset
\frak U (\hat{\hat{\frak g}})_n^m,\,\,\,\bigcap_{m \in \hf\Bbb Z}
\frak U (\hat{\hat{\frak g}})_n^m = 0,\,\,\,
\bigcup_{m \in \hf\Bbb Z}\frak U (\hat{\hat{\frak g}})_n^m =
\frak U (\hat{\hat{\frak g}})_n.$$
We take $\{\frak U (\hat{\hat{\frak g}})_n^m,m \in \hf\Bbb Z\}$ for
a fundamental neighborhood system of
$\frak U (\hat{\hat{\frak g}})_n$, and denote the corresponding
completion by
$\tilde{\frak U} (\hat{\hat{\frak g}})_n$. We let
$$\tilde{\frak U} (\hat{\hat{\frak g}}) = \bigoplus_{n\in \hf\Bbb Z}
  \tilde{\frak U} (\hat{\hat{\frak g}})_n.$$
Let $\langle {\bf k}-(k+h^{\vee}) \rangle$
be the two-sided ideal of the associative
superalgebra $\tilde{\frak U}(\hat{\hat{\frak g}})$ generated
by the element ${\bf k}-(k+h^{\vee})$. Denote
$\tilde{\frak U}(\hat{\hat{\frak g}})/\langle{\bf k}-(k+h^{\vee}) \rangle$
by $\tilde{\frak U}(\hat{\hat{\frak g}},k)$.

A $ \hat{\hat{\frak g}}$-module M is called {\em restricted} if for any
fixed $v \in M, x(n)v = 0 \,\,\,\mbox{for}\,\,\, n\gg 0$.
For example, $\hat{V}_k$
is a restricted module. The action of
$ \hat{\hat{\frak g}}$ on any restricted module can be extended to
$\tilde{\frak U} (\hat{\hat{\frak g}})$ naturally.

Let $\tilde{\frak U}(\hat{\hat{\frak g}},k)[[z,z^{-1}]]$
be the space of power series of $z,z^{-1}$ with coefficients in
$\tilde{\frak U}(\hat{\hat{\frak g}},k)$. An element
$b(z) = \sum_{n\in \Bbb Z}b(n)z^{-n-1}$ in
$\tilde{\frak U}(\hat{\hat{\frak g}},k)[[z,z^{-1}]]$
is called {\em regular} if every b(n) is homogeneous in
$\tilde{\frak U}(\hat{\hat{\frak g}},k)$ and
$\deg (b(n)) = -n + N_b$, where $N_b\in \hz$ is
a constant independent of $n$. We say that the regular
element is odd if $N_b\in \hf + \Bbb Z$, even
if $N_b\in \Bbb Z$. We define $\tilde{b}$ to be
1, if $b(z)$ is odd and 0 if $b(z)$ even. We denote by
$\tilde{\frak U}(\hat{\hat{\frak g}},k)\langle z \rangle$
the subspace linearly
spanned by the regular elements in
$\tilde{\frak U}(\hat{\hat{\frak g}},k)[[z,z^{-1}]]$.

Recall the $\hf \Bbb Z_{+}$-gradation
 $M_{k,0} = \oplus_{n\in\hf \Bbb Z_{+}}M_{k,0}(n)$.
Let $1 \in M_{k,0}(0)$ be the vacuum element of $M_{k,0}$.
Define $Y(1,z) = I \mid_{M_{k,0}}.$
We have
 $$M_{k,0}(0) = \Bbb C\cdot 1,\,\,\,
M_{k,0}(\hf) = \bar{\frak g}(-1)\cdot 1 \cong \bar{\frak g},\,\,\,
M_{k,0}(1) = \frak g (-1)\cdot 1 \cong \frak g.$$
For $a\in \frak g \subset M_{k,0}, \bar{a}\in\bar{\frak g}\subset M_{k,0},$
we define
$$a(z) = \sum_{n\in \Bbb Z} a(n) z ^{-n-1},
\bar{a}(z) = \sum_{n\in \Bbb Z} \bar{a}(n) z ^{-n-1}.$$
It is clear that $a(z),\bar{a}(z)\in \cpn.$
$a(z)$ is even while $\bar{a}(z)$ is odd.

\begin{definition}
For $b(z) = \sum_{n\in \Bbb Z}b(n)z^{-n-1}$,
and $a, \bar{a} \in M_{k,0},$ we define
\begin{eqnarray*} a(n)\bullet b(z) &=& Res_w(a(w)b(z)\iota_{w,z}(w-z)^n
- b(z)a(w)\iota_{z,w}(w-z)^n) \\
\bar{a}(n)\bullet b(z) &=& Res_w(\bar{a}(w)b(z)\iota_{w,z}(w-z)^n
-(-1)^{\tilde{b}} b(z)\bar{a}(w)\iota_{z,w}(w-z)^n)
\end{eqnarray*}  \label{def_bullet}
\end{definition}

By direct calculation, the following proposition which is
an anologue of Proposition 2.2.1 in [FZ] follows from
Definition \ref{def_bullet}.
\begin{proposition}
The definition above gives $\cpn$ the structure of a
$\hat{\hat{\frak g}}$-module, where ${\bf k}\in \hat{\hat{\frak g}}$
acts as $(k+h^{\vee}) \,I$. \label{prop_mod}
\end{proposition}

As a consequence of Definition \ref{def_bullet}, we have the following.
\begin{corollary}
For $a\in \frak g ,\bar{a}\in\bar{\frak g}$,
we have
$$ a(n)\bullet 1 = \left\{ \begin{array}{ll}
0,& n\geq 0
\\\frac{1}{(-n-1)!}(\frac{d}{dz})^{-n-1}a(z),& n < 0,
\end{array} \right.$$
$$ \bar{a}(n)\bullet 1 = \left\{ \begin{array}{ll}
0,& n\geq 0
\\\frac{1}{(-n-1)!}(\frac{d}{dz})^{-n-1}
\bar{a}(z),& n < 0.
\end{array} \right.$$ \label{cor_vac}
\end{corollary}

By Propositions \ref{prop_mod} and Corollary \ref{cor_vac},
we have a well-defined
homomorphism of $\hhg$-modules from $M_{k,0}$ to $\cpn$:
$$Y(\,\,,z):a_1 (-i_1)\cdots a_n (-i_n) \bar{b}_1(-j_1)
\cdots \bar{b}_m(-j_m)1 $$
$$\longmapsto
a_1 (-i_1)\bullet \cdots \bullet a_n (-i_n) \bullet \bar{b}_1(-j_1)
\bullet \cdots \bullet \bar{b}_m(-j_m)\bullet 1.$$
Moreover, $Y(a(-1)1,z) = a(z),
Y(\bar{a}(-1)1,z) = \bar{a}(z)$ for $a\in \frak g,
\bar{a} \in \bar{\frak g}$.

Since $M_{k,0}$ is a $\tilde{\frak U}(\hat{\hat{\frak g}},k)$-module,
we have a map from $\tilde{\frak U}(\hat{\hat{\frak g}},k)$
to $End ( M_{k,0})$. Now we have a series of maps
$$M_{k,0}\longrightarrow \cpn \subset
\tilde{\frak U}(\hat{\hat{\frak g}},k)[[z,z^{-1}]]
\longrightarrow End\ (M_{k,0})[[z,z^{-1}]].$$
We still denote the composition of these maps by
$$Y(\,\,,z):M_{k,0}\rightarrow End\ (M_{k,0})[[z,z^{-1}]].$$
For $b \in M_{k,0}$, we call $Y(b,z)$ the vertex operator of
$b$.

We use small letters
$a,b,c \cdots$ to denote the index among $1,2, \cdots, \dim \frak g.$
Choose a basis $\{u_a\}$ of $\frak g$  satisfying
$(u_a,u_b) = \hf \delta_{ab}$, $[u_a,u_b] = if_{abc}u_c,$
where $f_{abc}$ is anti-symmetric in $a,b,c$ and real valued
(These notations agree with those
in [KS]). Here and below we assume, as usual, summation
over repeated indices.

\begin{theorem}
$(M_{k,0}, 1,\omega,\tau, Y(\,\,,z))$ is an $N=1$ $ SVOA$
of rank $c_k$ provided that $ k \neq -h^{\vee}$,
 where
$$c_k = \frac{\dim \frak g}{2} + \frac{k \dim \frak g}{k + h^{\vee}},$$
$$\tau = \frac{2}{k+h^{\vee}}u_a (-1) \bar{u}_a (-1)1
+\frac{4i}{3(k+h^{\vee})^2}
f_{abc} \bar{u}_a (-1)\bar{u}_b (-1)\bar{u}_c (-1)1,$$
\begin{eqnarray*}
\omega &=& \frac{1}{k+h^{\vee}}\{u_a (-1) u_a (-1)1 \\
 &&+\bar{u_a}(-2)\bar{u_a}(-1)1\} + \frac{2i}{3(k+h^{\vee})^2}
f_{abc}\bar{u}_a (-1)\bar{u}_b (-1) u_c (-1)1.
\end{eqnarray*}
\label{thm_kt}
\end{theorem}
The fact that the components of the fields
$$Y(\tau,z) = \sum_{n\in \Bbb Z}G_{n+\frac{1}{2}}z^{-n-2},
\,\,\,\,\,\,Y(\omega,z)=\sum_{n\in \Bbb Z}L_nz^{-n-2},$$
satisfy the commutation relations of the Neveu-Schwarz algebra
with the central charge $c_k $ is ensured by Theorem 4 in [KT].
The rest of the proof of the above theorem
is similar to Theorem 2.4.1 in [FZ].

It follows from [KT] that
$[L_0, a(m)] = -m a(m),\,\,[L_0, \bar{a}(m)] = -(m+ \hf )\bar{ a}(m).$
Thus $ L_0 + d $, which commutes with all $a(m), \bar{a}(m)
\in \hhg$.
We call the generalized Casimir operator the element
$\Omega = 2(k + h^{\vee}) (L_0 + d).$
It follows from [KT] that
\begin{equation}
\Omega (v) = (\lambda + 2 \rho, \lambda )v
\label{eq_89}
\end{equation}
if $v$ is a singular vector of weight $\lambda$.

Let $J_{k,0}$ be the maximal proper submodule of $M_{k,0}$.
It is easy to see that if $ k\neq -h^{\vee}$ then
$1\not\in J_{k,0}, \tau \not \in J_{k,0}$
and hence the quotient $L_{k,0} =
M_{k,0}/J_{k,0}$ is also a $SVOA$.
To understand what $J_{k,0}$ is, we need to find the formulas
for singular vectors of $M_{k,0}$.
This is done in the Appendix (Sec.5).

\begin{remark}
One may construct the $N=2$ SVOA (i.e. the SVOA which admits
vertex operators whose Fourier components satisfy the $N =2 $
superconformal algebra) from the $(N =1) $ affine
Kac-Moody superalgebra [KS].
\end{remark}

\subsection{Rationality and fusion rules of
the SVOA $L_{k,0}$}

\begin{lemma}
The associative algebra $A(M_{k,0})$
is canonically isomorphic to $\frak U (\frak g ) $.
\end{lemma}

Proof: By the definition of $A(M_{k,0})$ and Lemma \ref{lemma_keylemma}
we have
$$ [c] * [a(-1)1] = [a(-1)c],$$
where $a\in \frak g,\,c \in M_{k,0}$. Hence
$$ [a_m (-1)1] *\cdots * [a_1 (-1)1] = [a_1 (-1)\cdots a_n (-1)1].$$
Therefore we have a homomorphism of associative algebras
\begin{equation}
 F: \frak U (\frak g)
\longrightarrow  A(M_{k,0}) \label{map}
\end{equation}
given by
$$ a_m \cdots a_1 \mapsto
[a_1 (-1)\cdots a_n (-1)1].$$
It is clear that
$$(a(-n-2) + a(-n-1))c = Res_z(Y(a(-1)1,z)\frac{z+1}{z^{n+2}}c),$$
$$\bar{b}(-n-1))c =
 Res_z(Y(\bar{b}(-1)1,z)\frac{1}{z^{n+1}}c).$$

By Lemma \ref{lemma_keylemma}, we have
$$ O'(M_{k,0})  \subset O(M_{k,0}), $$
where
$$ O'(M_{k,0}) = \{(a(-n-2) + a(-n-1))c , \,\,\,
\bar{b}(-n-1))c\,\,\,\mbox{for}\,\,\,n\geq 0\}.$$
Then it follows that
$$[a_1(-i_1 -1)\cdots a_m(-i_m -1)]= (-1)^{i_1+\cdots +i_m}
[a_1 (-1)\cdots a_n (-1)1]$$
for $i_1,\cdots i_m \geq 0.$
So F is an epimorphism. To show that F is indeed an isomorphism,
we still need to show that
\begin{equation}
 O'(M_{k,0}) = O(M_{k,0}).   \label{equn}
\end{equation}
However this is standard (see the proof of a similar
fact in Appendix of [W]). \,\,  $\Box$

\begin{lemma}
If $k$ is a positive integer, then the map (\ref{map}) induces
an isomorphism from $\frak U (\frak g )/\langle e_{\theta}^{k+1}\rangle $ onto
$A(L_{k,0})$, where $\langle e_{\theta}^{k+1}\rangle $ is the two-sided ideal
of $\frak U (\frak g )$ generated by $e_{\theta}^{k+1}$.
\label{lemma_ideal}
\end{lemma}
{\em Proof.} It follows from Theorem \ref{thm_reln}
in the Appendix that the SVOA $M_{k,0}$
is isomorphic to
$$M(\Lambda +h^{\vee} \Lambda_0)/
\langle f_i 1, i = 1,\cdots, l\rangle,$$ with $\Lambda$ given by
$\lambda_i = \Lambda(h_i) = 0,i = 1,\cdots,l,\,\,
\lambda_0 = \Lambda({\bf k}) = k.$ Then the SVOA $L_{k,0}$
is isomorphic to $M_{k,0}/\langle v_k\rangle$, where
$v_k$ is defined by Theorem \ref{theorem_singularvector2}.
By Remark \ref{rem_sing} and the identity (\ref{equn}), we see that
under the isomorphism (\ref{map}), $v_{\lambda_0}$ corresponds to
$e_{\theta}^{k+1} \in \,\,\frak U (\frak g )$. Hence the lemma follows
from Proposition \ref{proposition_associativity}.\,\, $\Box$

\begin{lemma}
If $x \in \frak g$, and $N \in \Bbb N$, then the algebra
$\frak U (\frak g )/\langle x^N \rangle $ is finite dimensional
and semisimple.
\label{lemma_fnty}
\end{lemma}
{\em Proof.} Let $G$ be the adjoint group of $\frak g$. Since
$G$ is generated by $exp (ad \,\,y), y \in \frak g$, the ideal
$\langle x^N \rangle $ is $G$-invariant, hence it contains
all elements $g(x)^N, g \in G$. Since $\frak g$ is simple,
it coincides with the linear span of the orbit $G(x)$, hence
$u_i^N \in \langle x^N \rangle $ for some basis $\{u_i\}$ of
$\frak g$. It follows that $dim \, \frak U (\frak g )/
\langle x^N \rangle \leq N^{dim \, \frak g}$.

Since  any finite-dimensional representation of
$\frak U (\frak g )$ is semisimple, it follows that any representation
of $\frak U (\frak g )/\langle x^N \rangle $ is semisimple.
Hence the latter algebra is semisimple. \,\,$\Box$

\begin{theorem}
For any positive integral $k$, the SVOA $L_{k,0}$ is rational.
Moreover, $L_{k,\lambda}$, for $\lambda \in {\frak h}^*$
dominant integrable with $\langle \lambda,\theta\rangle  \leq k$, are
precisely all the irreducible $L_{k,0}$-modules.
\label{thm_fz}
\end{theorem}
{\em Proof.} The second part of this theorem follows from
Theorem \ref{theorem_correspondence} and Lemma \ref{lemma_ideal}
because by Lemma \ref{lemma_fnty},
$L_{k,\lambda}$, for $\lambda \in {\frak h}^*$
dominant integrable with $\langle \lambda,\theta\rangle  \leq k$,
are all the irreducible modules of
$\frak U (\frak g )/\langle e_{\theta}^{k+1}\rangle $.
Any $L_{k,0}$-module $M$ is a restricted module over $\hhg$.
Hence any $\hhg$-submodule of $M$ is also an
$L_{k,0}$-submodule of $M$. To prove the complete
reducibility of any $L_{k,0}$-module, we only need to prove the
following.

\begin{lemma}
Given $\lambda,\,\mu \in P_{+}$ such that
$\langle \lambda, \theta \rangle \leq k,
\langle \mu, \theta \rangle \leq k,$ any
short exact sequence of $\hat{\hat{\frak g}}$-modules
$$0 \longrightarrow L_{k,\lambda}
    \stackrel{\iota}{\longrightarrow} M
    \stackrel{\pi}{\longrightarrow} L_{k,\mu}
    \longrightarrow 0$$
splits. \label{lemma_reducibility}
\end{lemma}
{\em Proof.} Let $Q_{+} = \Sigma_i \Bbb Z_{+} \alpha_i.$
First let us define a partial order in $P_{+}$
as follows: $ \lambda > \mu $ iff $ \lambda - \mu  \in Q_{+}$
and $\lambda \neq \mu $.
Without loss of generality, we may assume that $ \lambda \not > \mu $.
Otherwise we can apply the contragredient functor to the
short exact sequence to reverse it.
Let $v_{\mu}$ be the vacuum vector of $L_{k,\mu}$. Pick a
vector $v_{\mu }' \in M $ of weight $\mu$
such that $\pi ( v_{\mu}') = v_{\mu}$.
We claim that $v_{\mu}'$ is a singular vector of $M$,
i.e. $e_i v_{\mu}' =0 $ for any $i$. Indeed, if
$e_i v_{\mu}' \neq 0 $ for some $i$, then
$$\pi (e_i v_{\mu}') = e_i \pi (v_{\mu}') = e_i v_{\mu} = 0.$$
So
\begin{equation}
e_i v_{\mu}' = \iota (u) \label{eq_75}
\end{equation}
for some nonzero $u \in L_{k,\lambda}$, since the short sequence
is exact. Comparing the weights of both sides of equation
(\ref{eq_75}), we have $ \lambda - \beta = \mu + \alpha_i$
for nonzero $\alpha_i, \, \beta \in Q_{+}$.
It follows that $ \lambda = \mu +\alpha_i + \beta > \mu,$
which is a contradiction.

Denote by $M'$ the submodule of $M$ generated
by the singular vector $v_{\mu}'$. It suffices to show that
the module $M'$ is irreducible. But this follows in the
same way as in Chapter 11 of [K] by making use of
formula (\ref{eq_89}). \,\, $\Box$

Of course the Lie subalgebra
$$ \hat{\frak g} = \frak g \bigotimes \Bbb {C}[t,t^{-1}]
\bigoplus \Bbb {C}{\bf k} \bigoplus
\Bbb {C}d $$ of $\hhg$ is the usual affine
Kac-Moody algebra. As usual, we let
$$ \hat{\frak g}_{+} = \frak g \bigotimes t\Bbb C[t], $$
$$ \hat{\frak g}_{-} = \frak g \bigotimes t^{-1}\Bbb C[t^{-1}], $$
$$ \hat{\frak n}_{\pm} = \hat{\frak g}_{\pm}
\bigoplus  \bigoplus_{\alpha \in \Delta_{+}}\frak g_{\pm\alpha}.$$

Given $\lambda \in P_{+}$ and $k \in \Bbb Z_{+}$, consider the
irreducible $\frak g$-module $\bar{L}(\lambda)$ as a module over
$\hat{\frak g}_{+} \bigoplus \frak g \bigoplus \Bbb C {\bf k}
\bigoplus \Bbb C d$
by letting
$ (\hat{\frak g}_{+} \bigoplus \Bbb C d)\bar{L}(\lambda) = 0 $
and ${\bf k} = k I \mid_V.$ Let
$$\bar{M}_{k, \lambda} = \frak{U}(\hat{\frak g})
\bigotimes_{\frak{U}(\hat{\frak g}_{+}
\bigoplus \frak g \bigoplus \Bbb C {\bf k}
\bigoplus \Bbb C d)} \bar{L}(\lambda),$$
and $\bar{L}_{k,\lambda} =
\bar{M}_{k,\lambda} /\bar{J}_{k,\lambda}$, where $\bar{J}_{k,\lambda}$
is the unique maximal $\hat{\frak g}$-submodule of
$\bar{M}_{k,\lambda}$.

The associative algebra
of the VOA $\bar{L}_{k,0} $ was computed in [FZ].
Comparing with our results, we see that the
associative algebras $A(L_{k,0})$ and
$A(\bar{L}_{k,0})$ are the same. And so the irreducible modules of
the SVOA $L_{k,0}$ are canonically in 1-1 correspondence with those of
the VOA $\bar{L}_{k,0}$. One can calculate the fusion rules
using the $A(L_{k,0})$-modules similarly to Section 3.2
 in [FZ] and find
that the fusion rules for the modules of the SVOA $L_{k,0}$
are  canonically in 1-1 correspondence with those for
the VOA $\bar{L}_{k,0}$ (see the statements in Theorem 3.2.3
and Corollary 3.2.1 in [FZ]).

\section{Neveu-Schwarz SVOAs}
\setcounter{equation}{0}
\subsection{\em SVOA structure on $M_{c,0}$ and $L_{c,0}$}

Let us recall first that the Neveu-Schwarz
algebra is the Lie superalgebra
$$\frak {NS} = \bigoplus_{n \in \Bbb Z}  \Bbb{C}L_n
\bigoplus \bigoplus_{m \in \hf + \Bbb Z}  \Bbb{C}G_m
\bigoplus\Bbb{C}C$$
 with commutation relations ($m,n \in \Bbb Z $):
\begin{eqnarray}
[L_m,L_n] &=&(m-n)L_{m+n}+\delta_{m+n,0}\frac{m^3-m}{12}C, \nonumber
\end{eqnarray}
\begin{eqnarray}
[G_{m+\frac{1}{2}},L_n] & = &
(m+\frac{1}{2}-\frac{n}{2})G_{m+n+\frac{1}{2}}, \nonumber
\end{eqnarray}
\begin{eqnarray}
[G_{m+\frac{1}{2}},G_{n-\frac{1}{2}}]_{+} & = &
2L_{m+n}+\frac{1}{3}m(m+1)\delta_{m+n,0}C, \nonumber
\end{eqnarray}
\begin{eqnarray}
[L_m,C] = 0, \,\,\,\,\,
[G_{m+\frac{1}{2}},C] = 0.\nonumber \end{eqnarray}
The $\Bbb Z_2$-gradation is given by
$\tilde{L_n} = \tilde{C} = \bar{0},\, \tilde{G_{n}} = \bar{1}$
(so that the even part is the Virasoro algebra). Set
$$\frak {NS}_{\pm} = \bigoplus_{n \in \Bbb N}  \Bbb{C}L_{\pm n }
\bigoplus \bigoplus_{m \in \hf + \Bbb Z_{+}}  \Bbb{C}G_{\pm m}.$$
Given complex numbers $c$ and $h$, the Verma module $M_{c,h}$
over $\frak{NS}$ is the free $\frak{U} ({{\frak{NS}}_{-}})$-module
generated by 1, such that $\frak {NS}_{+}1 = 0, L_0 1 = h\cdot 1$
and $ C\cdot 1 =c\cdot 1.$
There exists a unique maximal
proper submodule $J_{c,h}$ of $M_{c,h}$. Denote
the quotient $M_{c,h} \,/ \,J_{c,h}$ by $L_{c,h}$. Recall that
$v \in M_{c,h}$ is called a singular vector if
$\frak {NS}_{+}1 = 0$ and $v$ is an eigenvector of $L_0$. For example,
 $G_{-\hf}1$ is a singular vector of $M_{c,0}$ for any $c$.
Denote $M_{c,0}\,/\langle G_{-\hf}1\rangle $ by $M_c$,
where $\langle G_{-\hf}1\rangle $ is
 the submodule of $M_{c,0}$ generated by
the singular vector $G_{-\hf}1$.
For simplicity we denote $L_{c,0}$ by $V_c$.

It is well known that
$$L_{-i_1}L_{-i_2} \cdots L_{-i_m}G_{-j_1}G_{-j_2}\cdots G_{-j_n},$$
for $i_1 \geq \cdots \geq i_m \geq 1,\,j_1 >\cdots > j_n \geq \hf,
i_1 \cdots i_m \in \Bbb N,$
and $ j_1 \cdots j_n \in \hf + \Bbb Z_{+}$
is a basis of $\frak U (\frak {NS}_{-})$.
There is a natural gradation on $M_{c,0}$, $ M_c$ and $V_c$ given by
the eigenspace decomposition of $L_0$:
$$\deg L_{-i_1}L_{-i_2} \cdots L_{-i_m}G_{-j_1}G_{-j_2}\cdots G_{-j_n}1
= i_1 + i_2 + \cdots + i_m + j_1 + \cdots + j_n.$$
We can define $\tilde{\frak U}(\frak {NS},c)$, the completion of
${\frak U}(\frak {NS})$, as in Section 2. The action of
$\frak {NS}$ on any restricted module of
$\frak {NS}$ can be extended to
$\tilde{\frak U} (\frak {NS})$ naturally. In particular,
$\tilde{\frak U} (\frak {NS},c)$ acts on $M_{c,h}$ and $M_c$.
We can also define the notions of even and odd regular elements
in $\tilde{\frak U}(\frak {NS},c)[[z,z^{-1}]]$. Denote by
$\tilde{\frak U}(\frak {NS},c)\langle z\rangle $
the linear span of regular elements
in $\tilde{\frak U}(\frak {NS},c)[[z,z^{-1}]]$.
Set
$$ L(z) = \sum_{n \in \Bbb Z} L_n z^{-n-2},$$
$$ G(z) = \sum_{n \in \Bbb Z} G_{n+\hf}z^{-n-2}.$$
Clearly $ L(z) $ is even while $ G(z) $ is odd.
For a regular element $b(z) \in \tilde{\frak U}(\frak {NS},c)[[z,z^{-1}]]$
we define
\begin{eqnarray}
\hspace{0.2 cm}
L_n \bullet b(z) &=& Res_w(L(w)b(z)\iota_{w,z}(w-z)^{n+1}
\label{eq_virvo}\\
 && -b(z)L(w)\iota_{z,w}(w-z)^{n+1}),
\nonumber
\end{eqnarray}
\begin{eqnarray}
\hspace{0.8 cm}
G_{n+\hf} \bullet b(z) &=& Res_w(G(w)b(z)\iota_{w,z}(w-z)^{n+1}
\label{eq_nsvo}\\
&&-(-1)^{\tilde{b}} b(z)G(w)\iota_{z,w}(w-z)^{n+1}),
\nonumber
\end{eqnarray}
where $\tilde{b}=0$ if $b(z)$ is even and $\tilde{b}=1$ if $b(z)$
odd.

{\bf Claim 6.1.}  (\ref{eq_virvo}) and  (\ref{eq_nsvo})
define a $\frak {NS}$-module
structure on $\tilde{\frak U}(\frak {NS},c)\langle z \rangle$
with central charge $c$.

{\bf Claim 6.2.}  $$\,\,L_n \bullet 1 = \left\{ \begin{array}{ll}
0,& n\geq -1 \\
\frac{1}{(-n-2)!}(\frac{d}{dz})^{-n-2}L(z),& n < -1,
\end{array} \right.$$
$$G_{n+\hf} \bullet 1 = \left\{ \begin{array}{ll}
0,& n \geq -1 \\
\frac{1}{(-n-2)!}(\frac{d}{dz})^{-n-2}G(z),& n < -1,
\end{array} \right.$$

{}From Claim 6.2 we have a well-defined homomorphism
of $\frak {NS}$-modules
\begin{equation}
Y(\,\,,z): M_{k,0} \rightarrow
\tilde{\frak U}(\frak {NS},c)\langle z\rangle  \label{eq_homom}
\end{equation}
$$L_{-i_1}L_{-i_2} \cdots L_{-i_m}G_{-j_1}G_{-j_2}\cdots G_{-j_n}1$$
$$\mapsto
L_{-i_1}\bullet L_{-i_2} \bullet\cdots \bullet L_{-i_m}\bullet
G_{-j_1}\bullet G_{-j_2}\bullet \cdots \bullet G_{-j_n}\bullet 1.$$
In particular, if we set $\tau = G_{-3/2}1,$
and $\omega = L_{-2}1,$
we have $Y(\tau,z) = G(z),$
and $Y(\omega,z) = L(z).$

Since $\tilde{\frak U}(\frak {NS},c)$ acts on $M_{c,h}$, we
have a map from $\tilde{\frak U}(\frak {NS},c)$
to $End\ (M_{k,0})$. Then (\ref{eq_homom}) induces a linear map
$$Y(\,\,,z): M_{k,0} \longrightarrow End\ (M_{k,0})[[z,z^{-1}]].$$

Thus we have

\begin{theorem}
$(M_c, 1,\tau, Y(\cdot,\,z))$ is an $N=1$ SVOA.
\end{theorem}

\subsection{\em Rationality and fusion rules of $V_{c_{p,q}}$ }

\begin{lemma}
There exists an isomorphism of associative algebras,
 $ F:A(M_c) \cong \Bbb{C}[x]$,
given by  $[\omega]^n \mapsto  x^n$,where $ \Bbb{C}[x]$
is the polynomial algebra on one generator $x$.
\label{lem_isom}
\end{lemma}

{\em Proof.}
 Set
$$ M_c = M_c^0 + M_c^1,$$
where $M_c^0\, ({\em resp.} M_c^1)$ is the even ({\em resp.} odd)
part of $ M_c  $.
By Lemma \ref{lemma_keylemma} we have
\begin{eqnarray}
(( L_{-m-3}+2L_{-m-2}+L_{-m-1})b) \label{eq_14}\\
\,\,\,\,\,\,\,\,\,\,\,\, = \mbox{Res}_z
(Y(\omega,z)\frac{(z+1)^2}{z^{2+n}}b)\in O(M_c), \nonumber
\end{eqnarray}
for every $m\geq 0, b \in M_c^0$ .
\begin{eqnarray}
(( G_{-n-1}+G_{-n})b) =  \mbox{Res}_z
(Y(\tau,z)\frac{(z+1)}{z^{1+n-\hf}}b)\in O(M_c),\label{eq_15}
\end{eqnarray}
for every $n\in \hf + \Bbb Z_{+}, b \in M_c^1$ .
It follows by induction that
\begin{eqnarray}
L_{-m} \sim (-1)^{m}((m-1)(L_{-2}+L_{-1})+L_0),
\label{eq_66}
\end{eqnarray}
for every $m \geq 1.$
\begin{eqnarray}
G_{-n} \sim (-1)^{n-\hf}G_{-\hf},\mbox{for every }\,n\in \hf + \Bbb Z_{+}.
\label{eq_666}
\end{eqnarray}
By Lemma \ref{lemma_keylemma}, we have
\begin{eqnarray}
[b] * [\omega] = [(L_{-2}+ L_{-1})b], b\in M_c^0
\label{eq_667}
\end{eqnarray}
Using (\ref{eq_14}) and (\ref{eq_15}), it is easy to show by induction
on  $m+n$ that
$$[L_{-i_1}L_{-i_2} \cdots L_{-i_m}G_{-j_1}G_{-j_2}\cdots G_{-j_{2n}}1]
= P ( [\omega] )$$
for some $ P(x)\in  \Bbb{C}[x]$.
Since the elements
$$L_{-i_1}L_{-i_2} \cdots L_{-i_m}G_{-j_1}G_{-j_2}\cdots G_{-j_{2n}}1$$
for $i_1 \geq \cdots \geq i_m \geq 1,\,j_1 >\cdots > j_{2n} \geq \hf,
i_1 \cdots i_m \in \Bbb N,\, j_1 \cdots j_{2n} \in \hf + \Bbb Z_{+},$
span $M_c$, the homomorphism of associative algebras
$$F:  \Bbb{C} [x]  \rightarrow  A(M_c)  $$
given by $ x^n \mapsto [\omega]^n $ is surjective. (This homomorphism is
well defined since $[\omega]$ is in the center of $A(M_{c})$.)

To prove that $F$ is also injective, it suffices to show that
$O(M_c)$ is the linear span of
the elements of the form (\ref{eq_14}) and (\ref{eq_15})
i.e.
$$O(M_c) = \{ ( L_{-n-3}+2L_{-n-2}+L_{-n-1})b, \,
(G_{-n-3/2} + G_{-n-3/2})b\,n\geq 0\,b \in M_c\}.$$
This can be proved in a standard way (for a proof
of a similar fact see Appendix of [W]). \,\, $\Box$


Set
$$ c_{p,q} = \frac{3}{2}(1-\frac{2(p-q)^2}{pq}), $$
$$ h_{p,q}^{r,s} = \frac{(sp-rq)^2 -(p-q)^2}{8pq}. $$
Whenever we mention $c_{p,q}$ again, we always assume
that $p\,, q\in \{2,3,4,\cdots \}, p-q\in 2 \Bbb Z$, and
that $(p-q)/2$ and $q$ are relatively prime to each other.
The submodule structure of a Verma module
over the Neveu-Schwarz algebra [A] is very similar to
that for the Virasoro algebras [FF].
{}From the results of [A], we have the following lemma which
is an analogue of the results in [FF] (also see Lemma 4.2 of [W]).
\begin{lemma}
\begin{enumerate}
\item [1)] $J_{c,0}$ is generated by the singular vector
$G_{- \hf }1$ if $c \neq c_{p,q}.$
\item [2)] $J_{c,0}$  is generated by two singular vectors if
$c = c_{p,q}$. One of them is $G_{- \hf }1$. The other, denoted by
$v_{p,q}$ has degree $\hf (p-1)(q-1).$
\end{enumerate}
\label{lem_sing}
\end{lemma}

{}From this lemma we immediately derive an analogue of
Corollary 4.1 in [W].

\begin{corollary}  If $c \neq c_{p,q}$, then $V_c$
is not rational.
\end{corollary}
{\em Proof.}
See proof of Corollary 4.1 in [W]. \,\,$\Box$

{ \bf  From now on, we always assume that ${\bf c = c_{p,q}}$}
{ \bf and that} ${\bf h^{r,s} = h^{r,s}_{p,q}}$.
It follows from Lemma \ref{lem_sing} that $V_{c} = M_{c} /
\langle v_{p,q} \rangle $, where
$\langle v_{p,q} \rangle$ denotes the submodule of $M_{c}$ generated
by $v_{p,q}$. Then we have

\begin{proposition}
One has:
$$A(V_c) \cong \Bbb C [x] / \langle F_{p,q}(x) \rangle,$$
where $\deg F_{p,q} = \frac{1}{4} (p-1)(q-1)$ if $p,q$ are odd;
$\deg F = \frac{1}{4} (p-1)(q-1) + \frac{1}{4}$ if $p,q$ are even.
\end{proposition}

{\em Proof.} If $p,q$ are odd, $v_{p,q}$ is an even element
of degree $\hf (p-1)(q-1)$ which corresponds to a polynomial
$F_{p,q}$ of degree $\frac{1}{4} (p-1)(q-1)$;
if $p,q$ are even, $v_{p,q}$ is an odd element
of degree $\hf (p-1)(q-1)$. From the definition of
the associative algebra $A(V_c)$, it is $G_{- \hf}v_{p,q}$ which
corresponds to $F_{p,q}$ of degree $\frac{1}{4} (p-1)(q-1) +\frac{1}{4}$.
\,\, $\Box$

We expect the following conjecture, which is an analogue
of Theorem 4.2 in [W], to be true.

\begin{conjecture} The vertex operator superalgebra
$V_{c_{p,q}}$ is rational. Moreover, the minimal series modules
$L_{c, h_{r,s}}, 0<r<p,  0<s<q, r-s \in 2 \Bbb Z$
are all the irreducible representations of $V_{c}$.
\label{thm_rat}
\end{conjecture}

\begin{remark}
If $p-q = 2$, $V_{c,h^{r,s}}$ is unitary. In this case we can prove
Theorem \ref{thm_rat} by using the well-known GKO construction [KW]
and Theorem \ref{thm_fz} (see [DMZ] for a similar proof in
the Virasoro algebra case). It follows that Conjecture \ref{thm_rat}
holds at least in the cases $p-q = 2$.

{}From the argument of Lemma \ref{lem_isom}, we see that
$$A(V_{c}) = H_0 (\frak S , V_c),$$
where $\frak S = \{ L_{-n-2} + 2L_{-n-1} + L_{-n},
G_{-n-1} + G_{-n}, n>0\} $ is a locally
nilpotent subalgebra of $\frak{NS}$.
This conjecture can probably be proved as in [W], by calculating
the coinvariants $H_0 (\frak S , V_c)$.
It is easy to see by Lemma \ref{lem_isom} that
$$A(L_{c,h^{r,s}}) = H_0 (\frak S , L_{c,h^{r,s}}).$$

Then by applying Theorem \ref{theorem_fusionrules},
and Proposition \ref{proposition_fusionrules}, we can
obtain the fusion rules for the $V_{c_{p,q}}$-modules
$L_{c, h_{r,s}}, 0<r<p,  0<s<q, r-s \in 2 \Bbb Z $
if the coinvariants $ H_0 (\frak S , L_{c,h^{r,s}}) $
are calculated.
\end{remark}

To support our conjecture, we present some examples.

{\em Example 1.} Consider the case $(p,q) = (5,3)$, $c_{5,3} = 7/10,$
$h^{1,1} = 0, h^{2,2} = 1/10.$
It is easy to check that the singular vector $v_{5,3}$ is
given by
$$v_{5,3} = 3L_{-4}1 + 10 L_{-2}^2 1-15 G_{-5/2}G_{-3/2}1.$$
Using (\ref{eq_66}), (\ref{eq_666}) and (\ref{eq_667}), we have
$$F_{5,3}(x) = 10 (x^2 - \frac{1}{10} x)$$ which gives the values
of $h^{1,1}$ and $ h^{2,2}.$

{\em Example 2.} Let $(p,q) = (8,2)$, $c_{8,2} = - \frac{21}{4},$
$h^{1,1} = 0, h^{3,1} = -\frac{1}{4}.$ This is a non-unitary case.
The singular vector $v_{8,2}$ is given by
$$v_{8,2} = 3G_{-7/2}1 - 4L_{-2}G_{-3/2}1.$$
Since $v_{8,2}$ is an odd element, we consider $G_{-\hf } v_{8,2}$
in order to get the polynomial $F_{8,2} (x)$. Using
(\ref{eq_66}), (\ref{eq_666}) and (\ref{eq_667}), we get
$$G_{-\hf } v_{8,2} \sim -8x(x + \frac{1}{4} )$$
which gives the values of $h^{1,1}$ and $ h^{3,1}.$

\section{SVOAs generated by free fermionic fields}
\setcounter{equation}{0}

The free fermionic fields are
\begin{eqnarray*}
\Phi^a (z) = \sum_{i \in \hf + \Bbb Z} \phi_{i}^a z^{-i-\hf},\,\,\,\,\,\,
(neutral)
\end{eqnarray*}
\begin{eqnarray*}
\Psi ^{a,\pm}(z) = \sum_{i \in \hf + \Bbb Z} \psi_{i}^{a,\pm}
z^{-i-\hf},\,\,\,\,\,\, (charged)
\end{eqnarray*}
with the following nontrival commutation relations
$$ [\phi_{i}^a, \phi_j ^b]_{+} = \delta_{a,b} \delta_{i,j}$$
$$ [\psi_{i}^{a,+},\psi_{j}^{b,-}]_{+}
   = \delta_{a,b} \delta_{i,j}, $$
where $a , b = 1, \cdots, l$.

It is easy to see that from a pair of charged free fermionic
fields $\Psi ^{\pm}(z)$ one can construct two neutral free fermionic fields
$\Phi (z)$ by letting
$$\Phi^1 (z) = \frac{1}{\sqrt{2}}(\Psi^{+} (z) + \Psi^{-} (z)),$$
$$\Phi^2 (z) = \frac{i}{\sqrt{2}}(\Psi^{+} (z) - \Psi^{-} (z)),$$
and vise versa.
Hence we only need to consider the SVOAs
generated by neutral free fermionic fields.
Let $F$ be the Fock space defined by
$\phi_{i > 0}^{a}|0\rangle = 0, a = 1,2, \cdots ,l.$ $F$ is
a SVOA with the Virasoro element
$\omega = \hf \sum_{a=1}^l \phi_{-3/2}^{a} \phi_{-1/2}^{a}1,$
and central charge $ c = \frac{l}{2}$.
Denote by $\frak a$ the Lie algebra linearly spanned
by $\{\phi^a_{n+\hf }, a= 1, \cdots, l, n \in \Bbb Z \}.$
$\frak U (\frak a)$ admits a natural gradation by letting
$\deg \phi^a_{n+\hf } = -n - \hf .$
Then, as in Subsec. 2.1, we can define the completion
$\tilde{\frak U}(\frak a)$ of $\frak U (\frak a)$.
Similarly, we can define the notion of an (even or odd)
{\em regular} vector in $\tilde{\frak U}(\frak a)[[z,z^{-1}]]$.
Let $\tilde{\frak U}(\frak a) \langle z \rangle$ be the linear
span of all regular vectors.
Define $Y(\phi^{a}_{- 1/2}1,z) = \Phi ^{a}(z).$
For any regular element $b(z) \in
\tilde{\frak U}(\frak a) \langle z \rangle$
we define an action
$$\phi^a_{n+\hf } \bullet b(z)
= Res_w (\Phi^a (w) b(z) \iota_{w,z} (w-z)^n
-(-1)^{\tilde{b}} b(z)\Phi^a (w) \iota_{z,w} (w-z)^n).$$
It is easy to see that $\phi^a_{n+\hf } \bullet b(z)$
is also a regular vector. Then we define
the vertex operator associated to any
$v = \phi^{a_1}_{{n_1}+\hf } \cdots \phi^{a_s}_{{n_s}+\hf } 1\in ~F$
by $$Y(v,z) =    \phi^{a_1}_{{n_1}+\hf } \bullet
\cdots \phi^{a_s}_{{n_s}+\hf }\bullet 1.$$
This definition turns out to be the same as that defined by the
normal ordering product [T].

\begin{theorem}
$F$ is a rational SVOA. Moreover, $F$ has a unique
irreducible representation, namely $F$ itself.
\end{theorem}
{\em Proof.} First we calculate the associative algebra $A(F)$.
By Lemma \ref{lemma_keylemma}, we have
$$\phi^{a}_{-\hf-n} v = Res_z (Y(\phi^{a},z)\frac{1}{z^{1+n}})
\in O(F),\,\, n \geq 0, \,\,v \in F.$$
Since
$$ \{\phi^{a}_{-\hf-n} v,\,\,1\leq a \leq l\,\, n \geq 0, \,\,v \in F\}
= \bigoplus_{n \in \hf \Bbb N} F_n,$$
we have $\bigoplus_{n \in \hf \Bbb N} F_n \subset O(F).$

On the other hand, it is easy to check by definition of $O(F)$ that
$O(F)\subset \bigoplus_{n \in \hf \Bbb N} F_n.$ Thus
$O(F) = \bigoplus_{n \in \hf \Bbb N} F_n,$ and so
$A(F) = F /O(F) \cong \Bbb C.$
Hence there exists a unique representation of the associative
algebra $A(F) \cong \Bbb C$, i.e $\Bbb C$ itself. By Theorem
\ref{theorem_correspondence}, there exists a unique representation
of $F$, i.e. $F$ itself.

The complete reducibility of modules of $F$ follows from
a similar argument to the proof of Lemma \ref{lemma_reducibility}. So
$F$ is rational.    $\Box$

\begin{remark}
The above SVOA $F$ is not an $N=1$ SVOA. To construct
the $N=1$ SVOAs one needs to add some bosonic fields. For example, one
can see that the Fock space of
one free bosonic field and one free neutral fermionic
field is an $N=1$ SVOA of rank $\frac{3}{2}$.
This is just the special case of the SVOA associated to the affine
Kac-Moody superalgebra corresponding to the 1-dimensional
Lie algebra $\frak g$.
\end{remark}

\section{\em Appendix: Singular vectors and defining relations for
the integrable representations of affine Kac-Moody superalgebras}

We continue using the notation on affine (super)algebras
introduced in Subsec.2.1.

Recall the triangular decomposition
$\hat{\frak g} = \hat{\frak n_{+}} \bigoplus
\hat{ \frak h} \bigoplus \hat{\frak n}_{-}$, where
$$\hat{\frak n}_{\pm} = \hat{\frak g}_{\pm}
\bigoplus (\bigoplus_{\alpha \in \Delta_{+}} \frak g_{\pm\alpha}).$$
Recall that for $\Lambda \in \hat{\frak h}^{*}$ we have the Verma
module $\bar{M}(\Lambda) =
\frak U (\hat{ \frak g})
\bigotimes_{\frak{U}(\hat{\frak h} + \hat{\frak n}_{+})} \Bbb C_{\Lambda}$
over $\hat{\frak g}$, where $\Bbb C_{\Lambda}$ is the 1-dimensional
$\frak{U}(\hat{\frak h} + \hat{\frak n}_{+})$-module defined by
$h \longmapsto \Lambda (h),\,\,\hat{\frak n}_{+}\longmapsto 0.$
Note that $\bar{M}_{k, \lambda}$, where $k = \Lambda({\bf k})$
and $\lambda = \Lambda \mid_{\frak h}$, is a quotient module of
$\bar{M}(\Lambda)$ and that $\bar{L}(\lambda)$ is the quotient
of $\bar{M}(\Lambda)$ by the maximal submodule.

Similarly we have the triangular decomposition
$\hhg = \hhn_{-} \bigoplus \frak h \bigoplus \hhn_{+},$
where $ \hhn_{\pm} = \hhg_{\pm} \bigoplus
\bigoplus_{\alpha\in \Delta_{+}}\frak g_{\pm \alpha},$
and for a given $\Lambda \in \hat{\frak h}^{*}$,
we define the Verma module $M(\Lambda)$ over $\hhg$,
so that ${M}_{k, \lambda}$ is a quotient of $M(\Lambda)$
and $L(\Lambda)$ is the irreducible quotient of $M(\Lambda)$.

Set $ e_0 = e_{-\theta}(1) , f_0 = e_{\theta}(-1), \mbox{and } \,
h_0 = \alpha_0 = \bf k - \theta .$
Given $\Lambda\in \hh^{*},$ let $\lambda_i =\Lambda(h_i)$.
Let $P_{+} = \{ \Lambda \in \hh^{*} \mid \Lambda (h_i) \in \Bbb Z_{+}\}$.

It is well known [K] that if $\Lambda\in P_{+},$
then $\{f_i^{\lambda_i +1}1, i = 0,1,\cdots, l\}$
are the singular vectors of $\bar{M} (\Lambda)$ which
generate the maximal proper submodule of $\bar{M}(\Lambda)$,
denoted by $\overline{\langle } f_i^{\lambda_i +1}1,
i = 0,1,\cdots, l \overline{ \rangle }$,
i.e. $$\bar{L} (\Lambda) \cong
\bar{M} (\Lambda )/\overline{\langle } f_i^{\lambda_i +1}1,
i = 0,1,\cdots, l \overline{ \rangle }$$
and that the $\hat{\frak g}$-modules $\bar{L}(\Lambda)$, $\Lambda
\in  P_{+},$ are all the unitary highest weight modules. For $\hhg$
the situation is similar. In more detail,
there exists a unique hermitian form $H(\cdot,\cdot)$ on
the Verma module $M(\Lambda)$ satisfying
$$H(1,1) = 1,$$
\begin{equation}
a(n)^{*} = \sigma(a)(-n),
\end{equation}
\begin{equation}
\bar{a}(n)^{*} = \overline{\sigma(a)}(-n-1),\label{eq_hermitian}
\end{equation}
where $*$ denotes the adjoint operator with respect to
the hermitian form $H(\cdot,\cdot)$.
Then $L(\Lambda) = M(\Lambda) / Ker\,\,H$. The $\hhg$-module is
called unitary if the form H on $L(\Lambda)$ is positive definite.
It is known that there exists a unique unitary
highest weight $\hhg$-module of level $h^{\vee}$, called
the minimal representation $F$ which is given by the
Fock space realization of the infinite dimensional
Clifford algebra (\ref{eq_clifford}) and as a $\hg$-module
is isomorphic to
$L( h^{\vee} d )$ [KT]. Furthermore [KT],
any unitary highest weight representation
of $\hhg$ is of the form $L(\Lambda +h^{\vee}d)$,
where $\Lambda \in P_{+}$, and that one has an isomorphism as
$\hg$-modules:
$$L(\Lambda +h^{\vee}d) \cong F \bigotimes \bar{L}(\Lambda).$$

{}From the construction of the minimal representation, we
also see that as $\hg$-modules
\begin{equation}
M(\Lambda +h^{\vee}d)
\cong  F \bigotimes \bar{M}(\Lambda) \label{eq_verma}
\end{equation}
It is clear that $\{f_i^{\lambda_i +1}1, i = 1,\cdots, l\}$
are the singular vectors of $M(\Lambda + h^{\vee}d )$.
By comparing the character formulas of both sides of
(\ref{eq_verma}),
we see that there also exists a unique singular vector of weight
$\Lambda + h^{\vee}d - (\lambda_0 + 1)\alpha_0$ in
$M(\Lambda +h^{\vee} d)$. To get an explicit
formula for this singular vector,
we need to introduce the following notion of special roots.

\begin{definition}
A root $\alpha$ in $\Delta_{+}$ is called
{\em special} if $\theta - \alpha$ is also a root.
\end{definition}

Denote by $\Bbb S$ the set of all special roots.
The following is an equivalent way to define the set $\Bbb S$:

\begin{remark}
The set $\Bbb S$ is also characterized by the property:
 $r_{\theta} (\alpha) = \alpha - \theta, $
if $\alpha \in \Bbb S$; $r_{\theta} (\alpha) = \alpha ,$
if $\alpha \in \Delta-(\Bbb S \cup \{\theta \}).$
Also we have:
\begin{equation}
\Bbb S \cup \{\theta \} = \{ \alpha \in \Delta_{+}|
r_{\theta} (\alpha) \in - \Delta_{+} \}.
\label{eq_53}
\end{equation}
\label{rem_s}
\end{remark}
\begin{lemma}
The number of special roots is $2(h^{\vee}-2)$.
\label{lemma_ww1}
\end{lemma}

{\em Proof.}
Choose the shortest $w \in W$ such that $w(\alpha_i) = \theta $
for some simple root $\alpha_i$ of $\frak g$.
It is not difficult to see that $l(w) = h^{\vee} -2$.
Pick a reduced expression $w= r_{i_1}\cdots r_{i_{h^{\vee} -2}}.$
We claim that the expression
\begin{equation}
r_{\theta} = r_{i_1}\cdots r_{i_{h^{\vee} -2}}r_i r_{i_{h^{\vee} -2}}
\cdots r_{i_1},
\label{eq_232}
\end{equation}
is reduced, i.e.
$l(r_{\theta}) = 2h^{\vee} -3.$
Indeed, first from the expression (\ref{eq_232}) of $r_{\theta}$
we see that $l(r_{\theta}) \leq 2h^{\vee} -3.$
Let $$\beta_1 = \alpha_{i_1}, \, \beta_2 = r_{i_1}(\alpha_{i_2}),
\cdots, \beta_{h^{\vee}-2} = r_{i_1} \cdots r_{i_{h^{\vee} -3}}
(\alpha_{i_{h^{\vee} -2}})$$
and let
$$\gamma_s := \theta - \beta_s = w r_i r_{i_{h^{\vee} -2}}
\cdots r_{i_{s+1}} (\alpha_{i_s}).$$
It is easy to see (using Remark 5.1) that
$\{ \beta_1, \cdots, \beta_{h^{\vee} -2},
\gamma_1, \cdots \gamma_{h^{\vee} -2} \} \subset \Bbb S $.
Hence $l(r_{\theta}) \geq 2h^{\vee} -3.$
Then the lemma follows since
$l(w) = \# \{ \alpha \in \Delta_{+}|w^{-1} (\alpha ) \in - \Delta_{+} \}$
for any $w \in W $.
\,\,$\Box$

\begin{remark}
It follows from the above proof that
\begin{equation}
\Bbb S = \{ \beta_1, \cdots, \beta_{h^{\vee} -2},
\gamma_1, \cdots \gamma_{h^{\vee} -2} \} .
\label{eq_321}
\end{equation}
The sum of two
elements from $\Bbb S \cup \{ \theta \}$ is a root if and only
if one of them is $ \beta_i$ and the other is $\gamma_i$.
\end{remark}

\begin{lemma}
Assume that $\delta_i \in \Delta_{+}-\{\theta\},
i= 1,2,\cdots,p$ satisfy
\begin{equation}
 \sum_{i=1}^{I}\delta_i = p \theta + \sum_{k}\eta_k
\,\,\,\mbox{for some}\,\,p \in \Bbb N \,\,
\mbox{and}\,\,\eta_k \in \Delta_{+}.
\label{eq_55}
\end{equation}
Then $I\geq 2p$, and at least $2p\,\,\delta_i's$
are contained in $\Bbb S$.
\label{lemma_ww2}
\end{lemma}

{Proof.} Assume that there are $q\,\,\delta_i's$ which
are contained in $\Bbb S$. By applying $r_{\theta}$
to both sides of the equation (\ref{eq_55}), it follows from
Remark \ref{rem_s} that
\begin{equation}
\sum_{i=1}^{I}\delta_i - q \theta
= -p \theta + \sum_{k}r_{\theta} (\eta_k ).
\label{eq_711}
\end{equation}
By subtracting (\ref{eq_711}) from (\ref{eq_55}), we have
\begin{equation}
(q-2p) \theta = \sum_{k}(\eta_k - r_{\theta} (\eta_k )).
\label{eq_712}
\end{equation}
By Remark (\ref{rem_s}), $\eta_k - r_{\theta} (\eta_k )$ is in
$Q_{+} = \sum_{\alpha \in \Delta_{+}} \Bbb Z_{+} \alpha $.
It follows that $q \geq 2p.$ \,\, $\Box$


\begin{theorem}
The element $ v_{\lambda_0} = \bar{e}_{-\theta}(1)\bar{e}_{-\theta}
\Pi_{\alpha \in \Bbb S}\bar{e}_{-\alpha}\cdot
e_{\theta}(-1)^{\lambda_0 +h^{\vee}+1}1$
is a singular vector in
$M(\Lambda + h^{\vee} d)$ of weight
$\Lambda + h^{\vee} d - (\lambda_0 + 1)\alpha_0$. (Here and further,
$\bar{e}$ stands for $\bar{e}(0)$, where $e \in \frak g.$)
\end{theorem}

A different arrangement of order
in $\Pi_{\alpha \in \Bbb S}\bar{e}_{-\alpha}$
only makes a difference in the sign of $v_{\lambda_0}$.
In the following proof,
for convenience we use $x(m+\frac{1}{2})$ to denote
$\bar{x}(m)$ in $\hat{\hat{\frak g}}, \, m \in \Bbb Z.$

{\em Proof.} It follows from  Lemma \ref{lemma_ww1}
that the weight of $v_{\lambda_0}$ is
$\mu =  \Lambda + h^{\vee} d - (\lambda_0 + 1)\alpha_0$.
To prove that $v_{\lambda_0}$ is a singular vector, it suffices to
prove that for every homogeneous element
$w \in M^s(\Lambda + h^{\vee} d )_{\mu}$,
we have $H(v_{\lambda_0},w) = 0$ and that $v_{\lambda_0} \neq 0.$
The latter statement will follow from another
formula for $v_{\lambda_0}$ given by
Theorem \ref{theorem_singularvector2}. Here we
prove the former one.

By (\ref{eq_hermitian}), it is enough to show that
\begin{equation}
H(e_{\theta}(-1)^{\lambda_0 + h^{\vee} + 1}1, \tilde{w}) = 0,
\label{eq_zero}
\end{equation}
where
\begin{equation}
\tilde{w} =
\Pi_{\alpha \in \Bbb S} e_{\alpha}(-\frac{1}{2})e_{\theta}(-\frac{3}{2})
e_{\theta}(-\frac{1}{2})\cdot w. \label{eq_w}
\end{equation}
Any homogeneous element $w$ in $M(\Lambda + h^{\vee} d )_{\mu}$
is of the form $$\Pi_{i=1}^{I}e_{\gamma _i}(-m_i)
\Pi_{j=1}^{J}e_{\theta}(-n_j)\Pi_{k=1}^{K} e_{-\eta _ k}(-l_k)\cdot 1,$$
where $\eta _ k \in \Delta_{+},\gamma _i \in \Delta_{+}-\{\theta\},\,\,\,
K,I,J \in \Bbb Z_{+}, l_{k} \in \frac{1}{2}\Bbb Z_{+},m_i ,n_j \in
\frac{1}{2}\Bbb N$,\,\,\, and
\begin{equation}
-\sum_{k} \eta_{k} + \sum_i \gamma_i + J\theta = (\lambda_0 +1)\theta,
\label{eq_res1}
\end{equation}
\begin{equation}
\sum_k l_k + \sum_i m_i + \sum_j n_j = \lambda_0 +1.
\label{eq_res2}
\end{equation}

Case 1) $n_j =\frac{1}{2}$ for some $j$.

Note that $e_{\theta}(-\frac{1}{2})$ commutes up
to a sign with elements of the form
$e_{\theta}(-n_j)$ or $e_{\gamma _i}(-m_i)$.
Since $e_{\theta}^2 (-\frac{1}{2})=0$, we have
$\tilde{w} = 0$ by (\ref{eq_w}).
(\ref{eq_zero}) is satisfied automatically.

Case 2) All $n_j \geq 1$.

On one hand, since all $m_i \geq \frac{1}{2}$, we have the inequality
\begin{equation}
\frac{I}{2} \leq \lambda_0 +1 - J
\label{eq_leq}
\end{equation}
since by (\ref{eq_res2}) we have
$$\frac{I}{2} = \sum_i \hf
\leq \sum_i m_i = (\lambda_0 +1) - \sum_j n_j - \sum_k l_k
\leq d +1 - \sum_j 1 = d +1 - J.  $$
And the equality in (\ref{eq_leq}) holds iff
\begin{equation}
m_i = \frac{1}{2}, \,\,\,
n_j = 1,\,\,\, l_k = 0.   \label{eq_12}
\end{equation}
for all $i,\,j,\,k.$

On the other hand, we rewrite (\ref{eq_res1}) as
$$ \sum_{i=1}^{I}\gamma_i = (\lambda_0 +1 - J) \theta + \sum_{k}\eta_k.$$
By  Lemma \ref{lemma_ww2},
\begin{equation}
I \geq 2(\lambda_0 +1 - J).   \label{eq_geq}
\end{equation}

Comparing (\ref{eq_leq}) and (\ref{eq_geq}),
we obtain that $ I = 2(\lambda_0 +1 - J)$ and then (\ref{eq_12}) holds.
Furthermore at least $2(\lambda_0 +1 - J)\,\,\gamma_i$'s are in
$\Bbb S$. Now we divide case 2) into two subcases:

Subcase 2.1) $ J < \lambda_0 +1$.

Then at least one $\gamma_{i_0}$
is in $\Bbb S$, i.e. $w$ can be expressed in the form
$e_{\gamma_{i_0}}(-\frac{1}{2})w'$. Then $\tilde{w} = 0$ since
$e_{\gamma_{i_0}}^2(-\frac{1}{2}) = 0$.

Subcase 2.2) $ J = \lambda_0 +1$.

Under this assumption we have $I=K=0$.
$w = e_{\theta}(-1)^{\lambda_0 +1}1.$
Then $ w'$ can be expressed in the form $e_{\theta}(-1)w''$
since $e_{\theta}(-1)$ commutes with
$\Pi_{\alpha \in \Bbb S} e_{\alpha}(-\frac{1}{2})e_{\theta}(-\frac{3}{2})
e_{\theta}(-\frac{1}{2}).$
By (\ref{eq_hermitian}),  we see that (\ref{eq_zero}) is equivalent to
$$ H(e_{-\theta}(1)e_{\theta}(-1)^{\lambda_0 + h^{\vee} + 1}1, w'') = 0.$$
However it is well known that
$e_{-\theta}(1)e_{\theta}(-1)^{\lambda_0 + h^{\vee} + 1}1 = 0$.     $\Box$

We can choose $e_{\pm \alpha} \in \frak g_{\pm \alpha},
\alpha \in \Bbb S \cup \{\theta \},$ in such a way that
\begin{equation}
[ e_{\alpha}, e_{-\alpha}] = - {\alpha} \label{eq_71}
\end{equation}
\begin{equation}
[e_{- \gamma_i}, e_{\theta}] = e_{\beta_i} \label{eq_72}
\end{equation}
\begin{equation}
[e_{-\beta_i}, e_{\theta}] = - e_{\gamma_i} \label{eq_73}
\end{equation}
Indeed, we pick $e_{- \gamma_i}$ and $e_{\theta}$ arbitrarily,
and define $e_{\beta_i}$ by the formula (\ref{eq_72}). Then
(\ref{eq_71}) fixes $e_{\gamma_i}$ and $e_{-\beta_i}$.
The formula (\ref{eq_73}) holds automatically since
$$( [ e_{-\beta_i}, e_{\theta}],e_{- \gamma_i})
=(e_{\theta}, [e_{- \gamma_i},e_{-\beta_i}])
= - (e_{\beta_i},e_{-\beta_i}) =1.$$

In the above notations, the singular vector $ v_{\lambda_0}$
can be written as (cf. (\ref{eq_321}))
$$ v_{\lambda_0} = \bar{e}_{-\theta}(1)\bar{e}_{-\theta}
\prod_{i = 1}^{h^{\vee}-2}(\bar{e}_{-\beta_i}\bar{e}_{-\gamma_i})\cdot
e_{\theta}(-1)^{\lambda_0 +h^{\vee} +1}1.$$
Note that $v_{\lambda_0}$ is independent of
the order of
$\prod_{i = 1}^{h^{\vee}-2}(\bar{e}_{-\beta_i}\bar{e}_{-\gamma_i})$.
Now we rewrite this formula of singular vectors in terms
of a PBW basis. We introduce a combinatorial symbol
$[m]_n = m(m-1)\cdots (m-n+1)$.

\begin{theorem}  One has:
\begin{eqnarray*}
\lefteqn{v_{\lambda_0} = \sum_{s=0}^{h^{\vee}-2}
\sum_{(i_1,\cdots,i_s)} (k+ h^{\vee })^{h^{\vee}-s-2}
[\lambda_0 + h^{\vee}+1]_{2(h^{\vee}-2)-s}\times}\\
& & \times \left( (k+ h^{\vee })
[\lambda_0 -s+3]_2 e_{\theta}(-1)^{\lambda_0 -s+1}\right.\\
& & \left.\,\,\,\,\,\,\,\, +[\lambda_0 -s+3]_3
e_{\theta}(-1)^{\lambda_0 -s}\bar{e}_{\theta}(-1)
\bar{h}_{\theta}(-1) \right. \\
& & \,\,\,\,\,\,\,\,
+\left.[\lambda_0 -s+3]_4 e_{\theta}(-1)^{\lambda_0 -s-1}
\bar{e}_{\theta}(-1)\bar{e}_{\theta}(-2) \right) Q_{i_1}\cdots Q_{i_s}
\cdot 1,
\end{eqnarray*}
where $Q_i = \bar{e}_{\beta_i}(-1)\bar{e}_{\gamma_i}(-1),$
and the sum $\sum_{(i_1,\cdots,i_s)}$ is taken over all
subsets of the set $ \{1,\cdots,h^{\vee}-2\}$.
\label{theorem_singularvector2}
\end{theorem}

{\em Proof.}  We assume that whenever some negative power
of $e_{\theta}(-1)$ appears in the following, the
corresponding monomial term is zero.

It is not hard to prove by induction that
\begin{eqnarray}
&& \hspace{2 cm}
\bar{e}_{-\beta_i}\bar{e}_{-\gamma_i}\cdot e_{\theta}(-1)^n
\label{eq_199}\\
&=& n(k+ h^{\vee })  e_{\theta}(-1)^{n-1}
+ n(n-1) e_{\theta}(-1)^{n-2}
\bar{e}_{\beta_i}(-1)\bar{e}_{\gamma_i}(-1)\nonumber \\
&& +e_{\theta}(-1)^n
\bar{e}_{-\beta_i}\bar{e}_{-\gamma_i}
 -n  e_{\theta}(-1)^{n-1}\bar{e}_{\beta_i}(-1)
\bar{e}_{-\beta_i} \nonumber \\
&& -n  e_{\theta}(-1)^{n-1}
\bar{e}_{\alpha_i}(-1)\bar{e}_{-\alpha_i}.\nonumber
\end{eqnarray}
It follows that
\begin{eqnarray}
&& \hspace{2 cm}
\bar{e}_{-\beta_i}\bar{e}_{-\gamma_i}\cdot e_{\theta}(-1)^n\cdot 1
\label{eq_175}\\
&=& n(k+ h^{\vee })  e_{\theta}(-1)^{n-1} \cdot 1
+ n(n-1) e_{\theta}(-1)^{n-2}
\bar{e}_{\beta_i}(-1)\bar{e}_{\gamma_i}(-1)\cdot 1.\nonumber
\end{eqnarray}
Using (\ref{eq_199}) and (\ref{eq_175}), we get by induction that
\begin{eqnarray*}
v' &:=&\Pi_{i = 1}^{h^{\vee}-2}\bar{e}_{-\beta_i}\bar{e}_{-\gamma_i}\cdot
e_{\theta}(-1)^{\lambda_0 +h^{\vee} +1}1\\
&= &\sum_{s=0}^{h^{\vee}-2}\sum_{(i_1,\cdots,i_s)}
 (k+ h^{\vee })^{h^{\vee}-s-2}
[\lambda_0 + h^{\vee}+1]_{\left(2(h^{\vee}-2)-s\right)} \times \\
&&\hspace{3 cm}\times
e_{\theta}(-1)^{\lambda_0 -s+3}Q_{i_1}\cdots Q_{i_s}\cdot 1.
\end{eqnarray*}
Since
$$[\bar{e}_{-\theta},Q_{i}] = 0$$
and
$$[\bar{e}_{-\theta},e_{\theta}(-1)^n] =
n(n-1)e_{\theta}(-1)^{n-2}\bar{e}_{-\theta}(-2)
 + ne_{\theta}(-1)^{n-1}\bar{h_{\theta}}(-1),$$
we have
\begin{eqnarray*}
\bar{e}_{-\theta}v' &=& \sum_{s=0}^{h^{\vee}-2}
\sum_{(i_1,\cdots,i_s)}(k+ h^{\vee })^{h^{\vee}-s-2}
[\lambda_0 + h^{\vee}+1]_{\left(2(h^{\vee}-2)-s\right)}\times \\
&&\times \left([\lambda_0 -s+3]_2
e_{\theta}(-1)^{\lambda_0 -s+1}\bar{e}_{-\theta}(-2)\right. \\
&&\hspace{0.5 cm}\left. +(\lambda_0 -s+3)e_{\theta}
(-1)^{\lambda_0 -s+2}\bar{h}_{\theta}(-1)\right)
Q_{i_1}\cdots Q_{i_s}\cdot v_0.
\end{eqnarray*}
Using another identity
$$[\bar{e}_{-\theta}(1),e_{\theta}(-1)^n] =
n(n-1)e_{\theta}(-1)^{n-2}\bar{e}_{-\theta}(-1)
+ ne_{\theta}(-1)^{n-1}\bar{h}_{\theta},$$
we get the desired formula.    $\Box$

\begin{remark}
It follows from Theorem \ref{theorem_singularvector2}
that $v_{\lambda_0} \neq 0$. Moreover
the only term which does not involve the odd
factors is a non-zero multiple of
$e_{\theta}(-1)^{\lambda_0 + 1}$. Therefore the submodule
of the Verma module $M(\Lambda +h^{\vee} d)$ generated
by $v_{\lambda_0}$ is again a Verma module. \label{rem_sing}
\end{remark}

\begin{theorem}
We have the following isomorphism
$$L(\Lambda +h^{\vee} d) \cong
M(\Lambda +h^{\vee} d)/
\langle v_{\lambda_0}, f_i^{\lambda_i +1}1, i = 1,\cdots, l\rangle,$$
where $\langle v_{\lambda_0}, f_i^{\lambda_i +1}1, i = 1,\cdots, l\rangle$
denotes
the submodule of $M(\Lambda +h^{\vee} d)$
generated by the singular vectors
$v_{\lambda_0}, f_i^{\lambda_i +1}1, i = 1,\cdots, l.$    \label{thm_reln}
\end{theorem}

{\em Proof}: Since the weights of $v_{\lambda_0},
f_i^{\lambda_i +1}1, i = 1,\cdots, l$
are $\Lambda + h^{\vee} d - (\lambda_i + 1)\alpha_i,
i = 0,1,\cdots, l$\,\,
respectively, we have the following isomorphism of $\hhg$-modules:
\begin{eqnarray*}
\lefteqn{\langle v_{\lambda_0}, f_i^{\lambda_i +1}1, i = 1,\cdots, l\rangle}\\
&=& \sum_{i=0}^{l} M(\Lambda + h^{\vee} d - (\lambda_i + 1)\alpha_i) \\
&=& \sum_{i=0}^{l} F \otimes \bar{M}(\Lambda - (\lambda_i + 1)\alpha_i)\\
&=&  F \otimes \overline{\langle } f_i^{\lambda_i +1}1,
i = 0,1,\cdots, l \overline{\rangle }.
\end{eqnarray*}

Therefore we have the following isomorphism of $\hhg$-modules:
\begin{eqnarray*}
\lefteqn{M(\Lambda +h^{\vee} d)/
\langle v_{\lambda_0}, f_i^{\lambda_i +1}1, i = 1,\cdots, l\rangle}\\
&=& \frac{ F \otimes \bar{M} (\Lambda)}
{ F \otimes \overline{\langle }f_i^{\lambda_i +1}1,
i = 0,1,\cdots, l \overline{\rangle }} \\
&=& F \otimes \bar{L}(\Lambda) \\
&=& L(\Lambda +h^{\vee} d). \,\,\,\,\,\,\,\,  \Box
\end{eqnarray*}

Department of Mathematics, MIT;

email addresses: kac, wqwang@math.mit.edu.


\begin{thebibliography}{ABCD99}

\bibitem[A]{[A]} A. Astashkevich,
{\em On the structure of Verma modules over
 Virasoro and Neveu-Schwarz algebras, preprint (1993) }


\bibitem[BPZ]{[BPZ]} A. Belavin, A. Polyakov and A. Zamolodchikov,
{ \em Infinite conformal symmetries in two-dimensional quantum
field theory, Nucl. Phys. B241(1984) 333-380 }

\bibitem[B]{[B]} R. Borcherds, {\em Vertex algebras,
Kac-Moody algebras, and the Monster,
Proc. Natl. Acad. Sci. USA. 83(1986)3068-3071}

\bibitem[DMZ]{[DMZ]} C. Dong, G. Mason and Y. Zhu, {\em Discrete
series of the Virasoro algebra and the moonshine module,
preprint}

\bibitem[FF]{[FF]} B.L. Feigin and D.B. Fuchs, {\em Verma modules
over the Virasoro algebra, Lect. Notes Math. 1060(1984)230-245 }

\bibitem[FHL]{[FHL]} I. B. Frenkel, Y. Huang and J. Lepowsky,
{\em On axiomatic approaches to vertex operator algebras
and modules, Mem. Amer. Math. Soc. vol 104, No. 494(1993)}

\bibitem[FLM]{[FLM]} I. B. Frenkel, J. Lepowsky and A. Meurman, {\em
Vertex operator algebras and the Monster, Academic Press, New York, (1988)}

\bibitem[FZ]{[FZ]} I. B. Frenkel and Y. Zhu, {\em Vertex operator algebras
associated to representations of affine and Virasoro algebra,
Duke Math. J., vol.66, No.1(1992)123-168}

\bibitem[K]{[K]} V. Kac, {\em Infinite dimensional Lie
algebras, third edition, Cambridge Univ. Press (1990)}

\bibitem[KS]{[KS]} Y. Kazama and H. Suzuki,
{\em New $N=2$ superconformal field theories and superstring
compactification, Nucl. Phys. B321(1989), 232-268}

\bibitem[KT]{[KT]} V. Kac and I. Todorov, {\em Superconformal
current algebras and their unitary representations,
Comm. Math. Phys. 102(1985) 337-347}

\bibitem[KW]{[KW]} V. Kac and M. Wakimoto, {\em Modular
invariant representations of infinite-dimensional Lie
algebras and superalgebras, Proc. Natl. Acad. Sci. USA,
vol.85(1988) 4956-4960}

\bibitem[KW]{[KW]} V. Kac and M. Wakimoto, {\em Unitarizable
highest weight representations of the Virasoro, Neveu-Schwarz
and Ramond algebras, Lect. notes Phys. 261(1986) 345-371}

\bibitem[L]{[L]} B.H. Lian, {\em On the classification of
simple vertex operator algebras, preprint (1992)}

\bibitem[T]{[T]} H. Tsukada, {\em Vertex operator superalgebra,
Comm. Alg. 18(7)(1990)2249-2274}

\bibitem[W]{[W]} W. Wang, {\em Rationality of Virasoro vertex operator
algebras, Duke Math. J., IMRN, vol.71, No.1(1993) 197-211}

\bibitem[Z]{[Z]} Y. Zhu, {\em Vertex operator algebras, elliptic
functions and modular forms, Ph.D. dissertation, Yale Univ.(1990)}
\end{thebibliography}
\end{document}